\documentclass[preprint,3p,authoryear,a4paper]{elsarticle}
\usepackage{graphicx}
\usepackage{epstopdf}
\usepackage{slashbox}
\usepackage{color}
\usepackage{dsfont}
\usepackage{amsmath}
\usepackage{amssymb}
\usepackage{amsfonts}
\usepackage{amsbsy}
\usepackage{caption}
\usepackage{subcaption}
\usepackage{amsthm}
\usepackage{array}
\usepackage{booktabs} 
\usepackage{verbatim}
\usepackage[english]{babel}
\usepackage{natbib}
\usepackage{ulem}
\usepackage{url}
\usepackage{multirow}
\usepackage{subcaption}
\usepackage{float}
\usepackage{graphics}

\renewcommand\appendix{\par
	\setcounter{section}{0}
	\setcounter{subsection}{0}
	\setcounter{table}{0}
	\setcounter{figure}{0}
	\gdef\thetable{\Alph{table}}
	\gdef\thefigure{\Alph{figure}}
	\gdef\thesection{\Alph{section}}
	\setcounter{section}{0}}

\numberwithin{equation}{section}

	\newcounter{arclist}

		\newcounter{arcenum}

\begin{document}

				\normalem
				
				\begin{frontmatter}
					
					\title{On the impact of outliers in loss reserving}
					
					\author[UMelb]{Benjamin Avanzi}
					\ead{b.avanzi@unimelb.edu.au}
					
					\author[UNSW]{Mark Lavender\corref{cor}}
					\ead{mark.lavender@live.com.au}
					
					\author[UNSW]{Greg Taylor}
					\ead{gregory.taylor@unsw.edu.au}
					
					\author[UNSW]{Bernard Wong}
					\ead{bernard.wong@unsw.edu.au}
					
					\cortext[cor]{Corresponding author. }
					
\address[UMelb]{Centre for Actuarial Studies, Department of Economics, University of Melbourne VIC 3010, Australia}
\address[UNSW]{School of Risk and Actuarial Studies, UNSW Australia Business School, UNSW Sydney NSW 2052, Australia}
					
%					\address[UdeM]{D\'epartement de Math\'ematiques et de Statistique, Universit\'e de Montr\'eal \\ Montr\'eal QC  H3T 1J4, Canada}
\begin{abstract}

		The sensitivity of loss reserving techniques to outliers in the data or deviations from model assumptions is a well known challenge. It has been shown that the popular chain-ladder reserving approach is at significant risk to such aberrant observations in that reserve estimates can be significantly shifted in the presence of even one outlier. As a consequence the chain-ladder reserving technique is non-robust. In this paper we investigate the sensitivity of reserves and mean squared errors of prediction under Mack's Model \citep*{Mac93}. This is done through the derivation of impact functions which are calculated by taking the first derivative of the relevant statistic of interest with respect to an observation. We also provide and discuss the impact functions for quantiles when total reserves are assumed to be lognormally distributed. Additionally, comparisons are made between the impact functions for individual accident year reserves under Mack's Model and the Bornhuetter-Ferguson methodology. It is shown that the impact of incremental claims on these statistics of interest varies widely throughout a loss triangle and is heavily dependent on other cells in the  triangle. 
Results are illustrated using data from a Belgian non-life insurer. 
		
\end{abstract}
\begin{keyword}
 Robust loss reserving \sep Chain-ladder \sep Mack's Model \sep Impact functions 
	
JEL codes: %http://www.aeaweb.org/journal/jel_class_system.php
%C44 \sep %Statistical Decision Theory; Operations Research
%C61 \sep %Optimization Techniques; Programming Models; Dynamic Analysis
%G24 \sep %Investment Banking; Venture Capital; Brokerage; Ratings and Ratings Agencies
G32 %\sep %Financing Policy; Financial Risk and Risk Management; Capital and Ownership Structure
%G35 %\sep %Payout Policy

MSC classes: 
%60G51 \sep % Processes with independent increments
%93E20 \sep % Optimal stochastic control
91G70 \sep 	%Statistical methods; risk measures [See also 62P05, 62P20] in Actuarial science and mathematical finance
%91G60 \sep 	%Numerical methods (including Monte Carlo methods) in Actuarial science and mathematical finance
62P05 %\sep 	%Applications of statistics to actuarial sciences and financial mathematics
%62H12 %\sep 	%Estimation in multivariate analysis
%91B30 %\sep % Risk theory, insurance
	
\end{keyword}

\end{frontmatter}
{\centering \large}

\setcounter{page}{1}
\section{Introduction}\label{intro}
\subsection{Motivation}
The classical loss reserving problem is interested in estimating existing (unpaid and/or unreported) claim liabilities,  based on past data. Traditional reserving techniques, widely used in practice, are applied on data at a certain level of aggregation (usually quarters or years), often presented in triangles. Parameters typically rely on only a few such data points - sometimes even only one. This makes those techniques very vulnerable to the presence of outliers.

This issue is well known, however, being able to quantify the \emph{specific} impact of each observation to certain statistics of interest provides greater information regarding the nature of the data at hand and will often provide insights regarding the techniques themselves. This is of particular importance when implementing and adjusting models. The objective of this paper is to provide a mathematically tractable approach to understanding how changes in each incremental claim in a loss triangle will impact certain statistics of interest. The presented impact functions are not designed to detect outliers but rather to deepen our understanding of how outliers may impact the results of a model. Impact functions coupled with statistically sound procedures to detect and treat abnormal observations will improve the robustness of reserving techniques and ultimately lead to more informed and reliable decisions. 

\citet*{VeTa08} calculate the impact of incremental claims on total reserve estimates under a range of models. They consider the traditional chain-ladder technique however the impacts are calculated numerically and hence no closed form equations are provided. Additionally, no consideration of individual accident years, mean squared errors (mse) or quantiles is given which constitutes a major contribution of this paper. 
\citet*{VeVaDh09} show that traditional chain-ladder reserve estimates are highly susceptible to even just one outlier and highlight that the impact on reserves may be positive or negative. See \citet*{VeVaDh09,VeDe11} for recent robust reserving techniques.

In this paper we rigorously investigate the impact that incremental observations have on reserve estimates, their variability, and their quantiles. Notably, we provide closed form equations for the first derivative of these statistics of interest under Mack's Model \citep*{Mac93}, which highlights numerous properties of this technique, including areas of a loss triangle where outliers are likely to have the greatest effect on results and hence where observations should be most heavily scrutinised. It appears that observations in the corners of a loss triangle have the potential to impact results most significantly. Additionally, we compare the impact of incremental observations on reserves under Mack's Model and the Bornhuetter Ferguson technique \citep*{BoFe72} which suggests that the latter approach is more robust.

These techniques may be applied in practice to identify areas of a given loss triangle that reserves are particularly sensitive to and hence where outliers, if present, may have a significant impact on results. These impact functions may also be used to compare reserve sensitivities under different techniques as we have done for Mack's Model and the Bornhuetter Ferguson approach. The impact that incremental observations are having in different loss triangles may be calculated using these impact functions and comparisons made between areas of sensitivity and properties of these different triangles. Through such a comparative study, trends may begin to emerge, making it easier to identify anomalous observations or even whole data sets with abnormal properties.

%\subsection{Structure of Paper} 
The paper is structured as follows. In Section \ref{S_Not} we define notation and briefly review the reserving techniques that will be considered. Impact functions are defined in Section \ref{impacts}. The following sections summarise the impact functions for certain statistics of interest and apply them to real data where 3D graphical representations are used to highlight their features. Section \ref{central} focuses on central estimates, Section \ref{S_MSE} on mean squared errors, and Section \ref{S_quantiles} on quantiles. The data used for the examples in this paper is presented and discussed  in Appendix \ref{data} and detailed proofs for the impact functions can be found in Appendix \ref{S:proof}. Section \ref{conclusion} concludes.

\section{Notation and Framework}\label{S_Not}

\subsection{Loss triangles}
The loss reserving problem is concerned with using currently available data to predict future claim amounts in a reliable manner.  The available data is often arranged in a loss triangle which provides a visual representation of the development of claims up to the current time as well as what is required to be predicted (see Figure \ref{losstriangle}). We denote by $X_{i,j}$ and $C_{i,j}$ the incremental and cumulative claims for accident year $i$ and development year $j$ respectively. Denote by $\mathbb{B}=\{X_{i,j}:i+j\leq I+1\}$ the past claims data. Let $R_i$ represent reserves for accident year $i$ and $R$ represent total reserves.

\begin{figure}[htb]
	\begin{center}
		\begin{tabular}{|c| c c c c c c|} \hline
			\textbf{$i$/$j$} & \textbf{1} & \textbf{2} & $\cdots$ & \textbf{$j$} &$\cdots$ & \textbf{$I$}  \\ \hline
			\textbf{1} & $X_{1,1}$ & $X_{1,2}$ & $\cdots$ & $X_{1,j}$ & $\cdots$ & $X_{1,J}$ \\ 
			\textbf{2} & $X_{2,1}$ & $X_{2,2}$ & $\cdots$ & $X_{2,j}$ & $\cdots$ & \\ 
			\vdots &\vdots &\vdots  &\vdots  &\vdots & &   \\ 
			\textbf{$i$} & $X_{i,1}$ & $X_{i,2}$ & $\cdots$ & $X_{i,j}$ & & \\ 
			\vdots &\vdots &\vdots & & & &    \\ 
			\textbf{I} & $X_{I,1}$ & & & & & \\ \hline
		\end{tabular}
	\end{center}
	\caption{Aggregate claims run-off triangle} 
	\label{losstriangle}
\end{figure}

\subsection{Chain-Ladder}
The traditional chain-ladder method is probably the most famous reserving technique. This approach hinges on the assumption that development factors 
$f_1,f_2,...,f_{I-1}$
exist, such that $
E[C_{i,j+1}|C_{i,j}]=f_jC_{i,j}$.
These development factors are unknown and estimated by 
\begin{eqnarray} \label{fhat}
\widehat{f}_j=\frac{\sum_{i=1}^{I-j}C_{i,j+1}}{\sum_{i=1}^{I-j}C_{i,j}},\ 1\leq j\leq I-1
.
\end{eqnarray}
Ultimate claims for accident year $i$ are then estimated by $
\widehat{C}_{i,I}=C_{i,I-i+1}\widehat{f}_{I-i+1}\cdots\widehat{f}_{I-1}$. 
From here accident year reserves and total reserves are subsequently estimated by \begin{equation}\label{Rihat}
\widehat{R}_i=\widehat{C}_{i,I}-C_{i,I-i+1}
\quad \text{and} \quad 
\widehat{R}=\sum_{i=1}^{I}\widehat{R}_i
\end{equation}

\subsection{Mack's Model}
Mack's Model \citep*{Mac93} is able to retain much of the simplicity of the deterministic chain-ladder whilst providing a formula for the mse of reserve estimates. %In particular, it remains distribution-free. 
The assumptions underlying classical chain-ladder reserves are the same for the first moment of reserves under Mack's Model. The mse of prediction for individual accident year reserves is given by \begin{eqnarray}\label{mse}
\text{mse}(\widehat{R}_i)=C_{i,I-i+1}\sum_{j=I-i+1}^{I-1}(f_{I-i+1}\cdots f_{j-1}\sigma^2_jf^2_{j+1}\cdots f^2_{I-1})+C_{i,I-i+1}^2(f_{I-i+1}\cdots f_{I-1}-\widehat{f}_{I-i+1}\cdots \widehat{f}_{I-1})^2
\end{eqnarray} 
The estimate for the mse of total reserves is given by \begin{equation}
\widehat{\text{mse}(\widehat{R})}=\sum_{i=2}^{I}\left\{\text{mse}(\widehat{R}_i)+\widehat{C}_{i,I}\left(\sum_{j=i+1}^{I}\widehat{C}_{j,I}\right)\sum_{k=I-i+1}^{I-1}\frac{2\widehat{\sigma}_k^2/\widehat{f}_k^2}{\sum_{n=1}^{I-k}C_{n,k}}\right\}
\end{equation}
where \begin{equation}
    \widehat{\sigma}_k^2=\frac{1}{I-k-1}\sum_{i=1}^{I-k}C_{i,k}(\frac{C_{i,k+1}}{C_{i,k}}-\widehat{f_{k}})^2,\ 1\leq k \leq I-2
\end{equation}
This paper is not conerned with tail-fitting, however, as outlined in \citep*{Mac93}, this approach leaves us without an estimator for $\sigma_{I-1}$ and we follow the same approach as in that paper to estimate $\sigma_{I-1}$ by requiring that $\frac{\widehat{\sigma}_{I-3}}{\widehat{\sigma}_{I-2}}=\frac{\widehat{\sigma}_{I-2}}{\widehat{\sigma}_{I-1}}$
    holds as long as $\widehat{\sigma}_{I-3}>\widehat{\sigma}_{I-2}$ leading to 
\begin{equation}
\widehat{\sigma}_{I-1}^2=\min(\frac{\widehat{\sigma}_{I-2}^4}{\widehat{\sigma}_{I-3}^2},\min(\widehat{\sigma}_{I-3}^2,\widehat{\sigma}_{I-2}^2))
\end{equation}

\subsection{Bornhuetter-Ferguson Reserves} \label{BF}
The Bornhuetter-Ferguson (BF) reserving methodology \citep*{BoFe72} is an opposing extreme to the chain-ladder (and Mack's Model) in that it uses prior estimates of ultimate claims and development patterns rather than inducing them from the available data to date. It can be considered a highly robust method as the presence of outliers will not influence reserve estimates. %This is the case if one is to use the pure BF method 
However in practice, chain-ladder development factors are often used to infer the development pattern when using the BF approach. It is this BF approach that we will consider (otherwise meaningful results will not present themselves). This means that the only difference between the two techniques is the estimate of ultimate claims for a given accident year. In particular, the BF method uses a prior (exogenously determined) estimate whereas the chain-ladder method uses the available data to estimate ultimate claims. Bornhuetter-Ferguson reserves are given by  \begin{eqnarray}
\widehat{R}_i^{BF}	& =	&\widehat{C}^{BF}_{i,I}-C_{i,I-i+1}= \widehat{\mu}_i-\widehat{\mu}_i\frac{1}{\widehat{f}_{I-i+1}\cdot...\cdot\widehat{f}_{I-1}} \quad \text{and } \quad \widehat{R}^{BF}=\sum_{i=1}^{I}\widehat{R}_i^{BF},
\end{eqnarray}
where $\widehat{\mu}_i$ represents the prior estimate of ultimate reserves for accident year $i$. 

\subsection{Using Impact Functions to Explore Reserve Sensitivities} \label{impacts}
In this section we present impact functions for numerous statistics of interest under the assumptions of Mack's Model. An impact function is able to highlight the sensitivity of a statistic of interest to a particular observation as well as pinpoint the marginal contribution of that observation to the final value of the statistic in some instances. This is done by taking the first derivative of the statistic with respect to the given observation. In our case we are interested in how an incremental claim $X_{k,j}$ may influence a given statistic $T$ such that the impact function is given by 
\begin{equation}
\text{IF}_{k,j}(T)=\frac{\partial T}{\partial X_{k,j}}.
\end{equation}
Further we have that if the statistic of interest $T$ is homogeneous of order one with respect to the $X_{k,j}$'s then 
\begin{equation}
T=\sum_{\{k+j\leq I-1\}}\frac{\partial T}{\partial X_{k,j}}X_{k,j}.
\end{equation}

The statistic, $T$, may represent reserves, the mse of reserve estimates or quantiles. It is interesting to investigate both the sign and magnitude of the impact functions to better understand the relationship a statistic has with an incremental claim. Furthermore, for those statistics that are homogeneous of order one, we may wish to see whether $\text{IF}_{k,j}(T)\cdot X_{k,j}$ is bounded or not. Boundedness will highlight that an outlying value of $X_{k,j}$ may only have a limited effect on T and hence this is a desirable property for robust estimators. If bounded, one should investigate the maximum value that $\text{IF}_{k,j}(T)\cdot X_{k,j}$ may take.

In the following subsections, we provide closed form equations to calculate the impact that each incremental claim is having on the aforementioned statistics in Mack's Model. We also provide the impact that incremental claims are having on reserves under the Bornhuetter-Ferguson methodology (see Section \ref{BF}).\\ The impact functions will be presented with the aid of an example using real data. This data is given in Appendix \ref{unidata}. By applying Mack's Model to this data set we calculate reserves for accident year 8 as \$226 403 952, total reserves as \$1 463 388 942, rmse for accident year 8 reserves as \$9 448 925 and the rmse for total reserves as \$45 480 914.

\section{Impact Functions on Central Estimates}\label{central}

\subsection{Individual Accident Years}\label{cenindivid}

\subsubsection{Mack's Model}\label{ifRi}

\begin{figure}[htb]
			\begin{center}
				\includegraphics[width=0.6\linewidth]{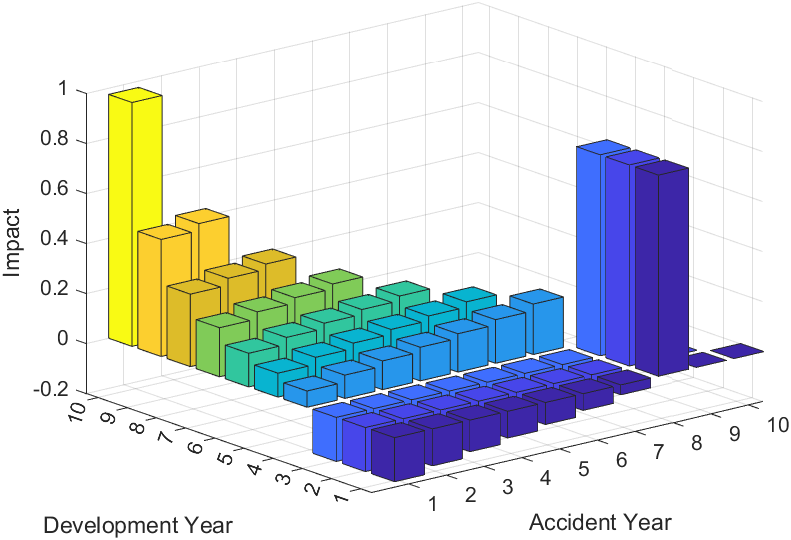}
				\caption{Illustration of $\text{IF}_{k,j}(\widehat{R}_8)$}
				\label{IFR8}
			\end{center} 
		\end{figure}

	The impact function for reserves of individual accident years ($\widehat{R}_i$) under Mack's Model is given by 
	\begin{eqnarray}
	\text{IF}_{k,j}(\widehat{R}_i)	& =	&\frac{\partial\widehat{R}_i}{\partial X_{k,j}} \\
	& =	& \begin{cases}\label{IFRi}
	0, &\text{if } k>i \\
	\frac{\widehat{R}_i}{C_{i,I-i+1}},& \text{if } k=i \\ 
	\widehat{C}_{i,I}\sum_{p=k}^{i-1}\left(\left(\frac{1}{\sum_{q=1}^{p}C_{q,I-p+1}}\right)\textbf{1}_{\{j\leq I-p+1\}}-\left(\frac{1}{\sum_{q=1}^{p}C_{q,I-p}}\right)\textbf{1}_{\{j\leq I-p\}}\right),& \text{if } k\leq i-1 
	\end{cases}\label{ifr} 
	\end{eqnarray}
	Where empty sums equal zero. See Appendix \ref{ifriproof} for proof. For $k=i$ this can be simplified further to a function only of future estimated development factors. This allows greater understanding of the impact of incremental claims for $k=i$ in that we can understand how these claims will affect  development factors by simply noting whether they will be represented in the numerator and/or denominator of equation \eqref{fhat}. 
	\begin{eqnarray}\label{kinf}
	\text{IF}_{i,j}(\widehat{R}_i)= \frac{\widehat{R}_i}{C_{i,I-i+1}}
	& =	& \frac{C_{i,I-i+1}\left(\prod_{s=I-i+1}^{I-1}\widehat{f}_s-1\right)}{C_{i,I-i+1}}= \prod_{s=I-i+1}^{I-1}\widehat{f}_s-1
	\end{eqnarray} 
An interesting point to note about the impact function for $\widehat{R}_i$ is that its value is heavily dependent on the position of the incremental claim in the loss triangle. In particular, three different cases for the accident year $k$ have been given in equation \eqref{IFRi} and furthermore the third case (i.e. $k\leq i-1$) includes a summation that is further dependent on the value of $k$ and two indicator functions that rely on the development period $j$. This dependence on position is a feature that is common to all impact functions provided in this paper. Notably, the impact of a positive change in any specific cell of a triangle may be positive in some other cells, negative in others, and the net effect on the reserve estimate may be positive or negative. This is an inherent effect of the reserving algorithm under consideration, and indicates how the results of that algorithm might be affected by an outlier in the original cell. Table \ref{egifri} provides the impact of each incremental claim on accident year 8 reserves. A 3D graphical representation of these impacts is given in Figure \ref{IFR8}.
	\begin{table}[htb]
		\begin{center}
			\resizebox{\linewidth}{!}{%
				\begin{tabular}{|c| c c c c c c c c c c|} \hline
					\textbf{$i$/$j$} & \textbf{1} & \textbf{2} & \textbf{3} &  \textbf{4} & \textbf{5} &  \textbf{6} & \textbf{7} & \textbf{8} & \textbf{9} & \textbf{10} \\ \hline
					\textbf{1} &-0.1762&	-0.1762&	-0.1762&	0.0649&	0.0955&	0.1346&	0.1961&	0.2899&	0.4679&	0.9748 \\ 
					\textbf{2} & -0.1479&	-0.1479&	-0.1479&	0.0932&	0.1238&	0.1628&	0.2244	&0.3182&0.4962&	    \\ 
					\textbf{3} & -0.1262&	-0.1262&	-0.1262&	0.1149&	0.1455&	0.1845&	0.2461&	0.3398&	& \\ 
					\textbf{4} & -0.1067&	-0.1067	&-0.1067&	0.1344&	0.1650&	0.2040&	0.2656& &  &   \\ 
					\textbf{5} & -0.0878&	-0.0878	&-0.0878&	0.1533	&0.1839	&0.2229&  &  &  &    \\  
					\textbf{6} & -0.0667&	-0.0667	&-0.0667&	0.1744&	0.2050 &  &  &  &  &   \\ 
					\textbf{7} & -0.0394&	-0.0394&	-0.0394&	0.2017&  & & & & &  \\ 
					\textbf{8} &0.8037&	0.8037&	0.8037 &&  & & & & &  \\ 
					\textbf{9} &0&	0  &  &  &  & & & & &  \\ 
					\textbf{10} & 0 &  &  &  &  & & & & &  \\ \hline
				\end{tabular}}
			\end{center}
			\caption{$\text{IF}_{k,j}(\widehat{R}_8)$} 
			\label{egifri}
		\end{table}
	\noindent The first case in equation \eqref{IFRi} is represented to the right of accident year 8 where incremental claims from accident years greater than the year of interest have no impact on reserves. The row of columns with equal height for accident year 8 corresponds to the second case of equation \eqref{IFRi} where incremental claims are having an equal and positive effect on the reserve estimate. Now, the area in the upper left of the loss triangle (i.e. between accident year 7 and development year 3) represents an area where all incremental claims are having a negative impact on reserves. More specifically, $ \text{IF}_{k,j}(\widehat{R}_i)\leq 0, \text{ for all } k\leq i-1 \text{ and } j\leq I-i+1
$. For $j>I-i+1$ (and $k\leq i-1$), the situation is somewhat murkier and we have the result that \begin{eqnarray} \text{IF}_{k,j}(\widehat{R}_i)>0, \text{ if }\sum_{p=k}^{\min\{i-1,I-j+1\}}\left(\left(\frac{1}{\sum_{q=1}^{p}C_{q,I-p}}\right)+\left(\frac{1}{\sum_{q=1}^{p}C_{q,I-p+1}}\right)\right)<\left(\frac{1}{\sum_{q=1}^{I-j+1}C_{q,j}}\right)
	\end{eqnarray} Where $\mathbb{D} = \{1, ..., i-1\}$. This inequality is readily calculable from the original loss triangle and we have found that in most instances we have considered it holds true. Figure \ref{IFR8} represents when this inequality holds as we see that for $j>I-i+1=3$, all impacts are positive.
	
	Additionally, note that for any choice of development period $j$ the impact is increasing with accident year $k$ throughout the loss triangle. Now we focus on the diagonals when $j>I-i+1$. For the most recent diagonal (i.e. $k+j=I+1$) we have that  $  \text{IF}_{k,j}(\widehat{R}_i)>\text{IF}_{k+1,j-1}(\widehat{R}_i) \Longleftrightarrow \sum_{q=1}^{k}X_{q,j}<C_{k+1,j-1}
	$. 
	This says that the impact will be increasing as we move up the most recent diagonal (from accident year $k+1$ and development year $j-1$ to accident year $k$ and development $j$) if the sum of incremental claims in column $j$ is less than the cumulative claims up to development year $j-1$ for accident year $k+1$. It is likely that this will hold for situations when incremental claims in later development periods are usually less than those in earlier periods and hence the column sums in these development years can be expected to be less than cumulative claims for the following accident year. Additionally, if this decreasing development pattern is present it will be more likely for this inequality to hold at later development periods than earlier ones. For the other diagonals we have that \begin{eqnarray}\text{IF}_{k,j}(\widehat{R}_i)>\text{IF}_{k+1,j-1}(\widehat{R}_i) \Longleftrightarrow\frac{1}{\sum_{q=1}^{k}C_{q,I-k+1}}-\frac{1}{\sum_{q=1}^{k}C_{q,I-k}}>\frac{1}{\sum_{q=1}^{k+l}C_{q,I-k+1-l}}-\frac{1}{\sum_{q=1}^{k+l-1}C_{q,I-k+1-l}},
	\end{eqnarray}
	where $l$ represents the diagonal that is being evaluated such that for the second most recent diagonal $l=2$, for the third most recent diagonal $l=3$ and so on. Note that in most examples that we have considered these inequalities hold and as a result we see the impact increasing for incremental claims as we move towards the top right hand corner of the loss triangle.
	
	The final property that we have derived is that for fixed accident year $k$, $\text{IF}_{k,j}(\widehat{R}_i)$ is increasing with $j$ for $j\geq I-i+1$. The proofs for these properties are given in Appendix \ref{ifriprops}.

\subsubsection{Bornhuetter-Ferguson}
\begin{figure}[htb]
			\begin{center}
				\includegraphics[width=0.6\linewidth]{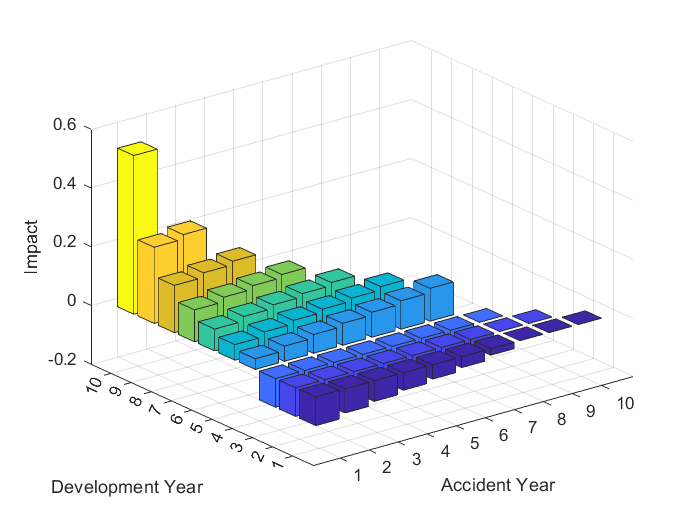}
				\caption{Illustration of $\text{IF}_{k,j}(\widehat{R}^{BF}_8)$}
				\label{IFBF}
			\end{center} 
		\end{figure}

	The impact function for BF individual accident year reserves is given by \begin{equation}\
		\text{IF}_{k,j}(\widehat{R}^{BF}_i)=\begin{cases}
		0 & \text{if } k\geq i\\
		\widehat{\mu}_i\frac{\sum_{p=k}^{i-1}\left(\left(\frac{1}{\sum_{q=1}^{p}C_{q,I-p+1}}\right)\textbf{1}_{\{j\leq I-p+1\}}-\left(\frac{1}{\sum_{q=1}^{p}C_{q,I-p}}\right)\textbf{1}_{\{j\leq I-p\}}\right)}{(\widehat{f}_{I-i+1}\cdot...\cdot\widehat{f}_{I-1})} & \text{if } k<i
		\end{cases}\label{BFimp}
		\end{equation}
Note that this result is a function of the assumption that $\widehat{\mu}_i=\widehat{C}_{i,I}$ and this is set before the calculation and unchanged in the calculation of the derivative otherwise we would receive the same results as under Mack's Model. We will discuss some  of the interesting results for $\text{IF}_{k,j}(\widehat{R}^{BF}_i)$ with the aid of Figure \ref{IFBF} which shows the impact function for Bornhuetter-Ferguson accident year 8 reserves under the assumption that $\widehat{\mu}_i=\widehat{C}_{i,I}$. The results for $\text{IF}_{k,j}(\widehat{R}_i^{BF})$ differ from the corresponding impact function for the chain-ladder reserves $(\text{IF}_{k,j}(\widehat{R}_i))$ in two major ways. Firstly, incremental claims in the same accident year as the reserve under inspection have no impact on that reserve in the BF case whereas they do in the CL case.  This is shown by zero values for each accident year greater than or equal to 8 in Figure \ref{IFBF}. %\BAc{really? double check - I agree with text, but the figure seems to have swapped axes?}This adds to the argument that the BF method is more robust than the CL. 
 Secondly, for the case when $k<i$, under the CL approach the $\widehat{\mu}_i$ is instead replaced by $\widehat{C}_{i,I}$ and there is no denominator term (i.e. $(\widehat{f}_{I-i+1}\cdot...\cdot\widehat{f}_{I-1})$ is not there). If the prior estimate of ultimate claims $\widehat{\mu}_i$ is less than or reasonably close to the CL estimate of ultimate claims $\widehat{C}_{i,I}$ then the impact of incremental claims under the BF method is less than the corresponding impact under the CL approach as it is divided by the factor $(\widehat{f}_{I-i+1}\cdot...\cdot\widehat{f}_{I-1})$. Of course this assertion will be dependent on the difference between $\widehat{\mu}_i$ and $\widehat{C}_{i,I}$ particularly because $(\widehat{f}_{I-i+1}\cdot...\cdot\widehat{f}_{I-1})$ may be only slightly greater than 1 in some instances.
 Further, $(\widehat{f}_{I-i+1}\cdot...\cdot\widehat{f}_{I-1})$ will be increasing with accident year $i$ such that the relative difference between the impacts under the BF and CL will increase with $i$.
\noindent Apart from the aforementioned changes in magnitude (and the change for $k=i$), the trends that we observe for this impact function will be similar to what was described in Section \ref{ifRi} for $\text{IF}_{k,j}(\widehat{R}_i)$. The proof for this impact function as well as a formal statement regarding the relationship between $\text{IF}_{k,j}(\widehat{R}_i)$ and $\text{IF}_{k,j}(\widehat{R}^{BF}_i)$ is given in Appendix \ref{bfproof}. 
	\subsection{Total Reserves}
	\subsubsection{Mack's Model}
	We have that $
	\widehat{R}=\sum_{i=1}^{I}\widehat{R}_i
$, such that the impact function for total reserves is simply given by 
	\begin{eqnarray}
	\text{IF}_{k,j}(\widehat{R})=\frac{\partial}{\partial X_{k,j}}\sum_{i=1}^{I}\widehat{R}_i=\sum_{i=1}^{I}\text{IF}_{k,j}(\widehat{R}_i) \label{totinf}
	\end{eqnarray}
	Again we use the aid of a diagram to illustrate the main properties of the impact function. 
		\begin{figure}[htb]
			\begin{center}
				\includegraphics[width=0.6\linewidth]{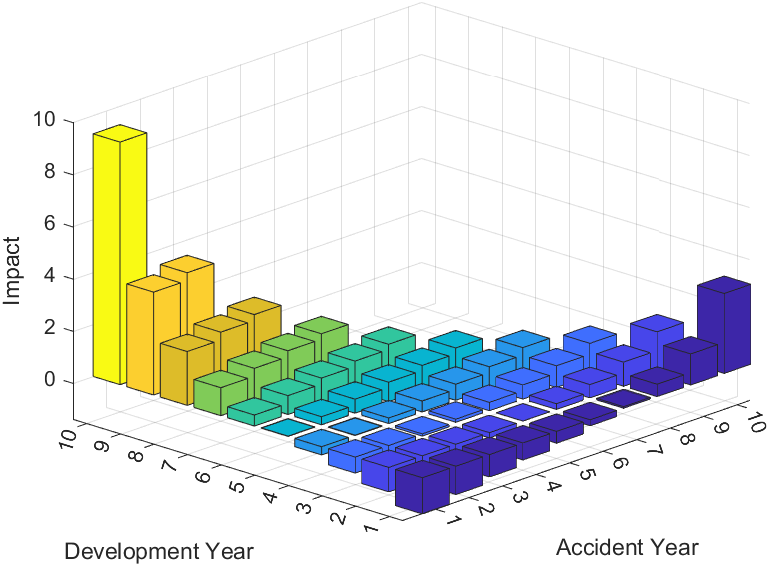}
				\caption{Illustration of $\text{IF}_{k,j}(\widehat{R})$}
				\label{VeVaDh091IFR}
			\end{center} 
		\end{figure}
\noindent The observation in the upper left corner of a loss triangle ($X_{1,1}$), has a negative impact on reserves for each accident year in every case. This cornerpoint is shown as the closest observation in Figure \ref{VeVaDh091IFR} and is having the largest negative impact on total reserve estimates (-1.3875). \\ \\  The impacts towards the latest development periods are also significant however they are positive. Importantly, this positive impact usually increases for each accident year as we move towards the upper right corner observation and hence this cornerpoint ($X_{1,I}$) will likely have a large positive impact. In our example, this observation ($X_{1,10}$) has the largest impact on total reserves (9.3050). The increasing pattern towards the top right corner can be understood by noting that the impact function for each accident year is increasing with $j$ for the same $k$ when $j\geq I-i+1$. However the result for final reserves is somewhat dependent on the inequalities as given in the Section \ref{ifRi} regarding the diagonals for $j\geq I-i+1$. %We have noted these inequalities usually hold.

The result for $X_{1,I}$ can be further understood by noting that any positive increase in this observation leads to a greater estimate of $f_{I-1}$ without a decrease in another estimated development factor. Hence final reserve estimates will be increased as this development factor is used for forecasting final cumulative claims for every other accident year.

Next, the bottom left corner observation ($X_{I,1}$) is the only observation currently available for the final accident year. The impact this observation has on final reserves is given solely by Equation \eqref{kinf}. Importantly, this value is greatest when considering observations in the first column as there are more development factors being multiplied than when $j>1$. Furthermore, in the other accident years (i.e. $k\neq I$), observations will be impacting estimated development factors $\widehat{f}_s$ such that one development factor will be increased and the other decreased as observations are altered. This is true except for the first column where the impact will only be felt for $\widehat{f}_1$ and it will be negative, and the last column where only $\widehat{f}_{I-1}$ will be impacted and the impact will be positive.

For the first column of observations, the impact will be negative or zero for each accident year except when $k=i$. We see that those observations around $X_{1,1}$ also often have negative impacts as they are encapsulated in the set $k\leq i-1$ and $j\leq I-i+1$ where their impact is negative for a larger number of accident years than other observations.

An additional interesting result is that for constant $k$ the impact is increasing with $j$ and for constant $j$ the impact is increasing with $k$ throughout this triangle. The impact is also increasing as we move along diagonals towards the top right corner for all $j\geq4$. These results can be understood by noting similar properties in the impact function for individual accident year reserves. % since these impacts are simply a summation of the associated individual impacts.

Similar results as to what has been stated here are mentioned on a heuristic basis in \citet*{VeTa08}. Notably, \citet*{VeTa08} highlight that impact functions can be used to evaluate the robustness of models and in turn compare and refine models based on robustness. This work provides mathematical justification for these conclusions and allows the impact of each observation to be traced precisely. These impact functions also provide insight into how adjustment of outlying points will affect results. 

\begin{figure}[htb]
		\begin{center}
			\includegraphics[width=0.6\linewidth]{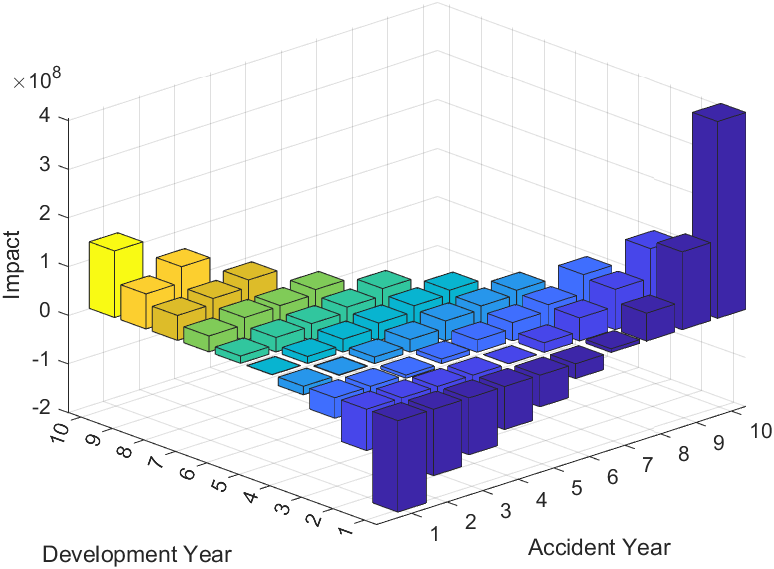}
			\caption{Illustration of $\text{IF}_{k,j}(\widehat{R})\cdot X_{k,j}$}
			\label{IFRmarg}
		\end{center} 
	\end{figure}

Note that the value of $\text{IF}_{k,j}(\widehat{R})$ and $\text{IF}_{k,j}(\widehat{R}_i)$ is independent of $X_{k,j}$ for $k+j=I+1$ (i.e. the last diagonal of the loss triangle). The proof for this result is given in Appendix \ref{independence}. A further result is that $\widehat{R}_i$ and $\widehat{R}$ are homogeneous of order 1 such that \begin{equation}
	\widehat{R}_i=\sum_{k+j\leq I+1}\text{IF}_{k,j}(\widehat{R}_i)\cdot X_{k,j} \qquad\text{and}\qquad \widehat{R}=\sum_{k+j\leq I+1}\text{IF}_{k,j}.(\widehat{R})\cdot X_{k,j}
	\end{equation}
	The proof for homogeneity of order 1 is given in Appendix \ref{homogeneity}. This allows us to find the marginal contribution of each incremental claim to reserves given by $\text{IF}_{k,j}(\widehat{R})\cdot X_{k,j}$. The 3D graph for these marginal contributions to total reserves is given in Figure \ref{IFRmarg} and we note that the result is somewhat different than when considering $\text{IF}_{k,j}(\widehat{R})$. This highlights how the magnitude of the incremental claim itself can impact the contribution it makes to reserves. In particular, we note that  the magnitude of incremental claims significantly decreases in later development periods and this is reflected in the graph. Hence, this analysis allows us to identify influential observations within a loss triangle. 
	
\subsubsection{Bornhuetter-Ferguson}
\begin{figure}[htb]
		\begin{center}
			\includegraphics[width=0.6\linewidth]{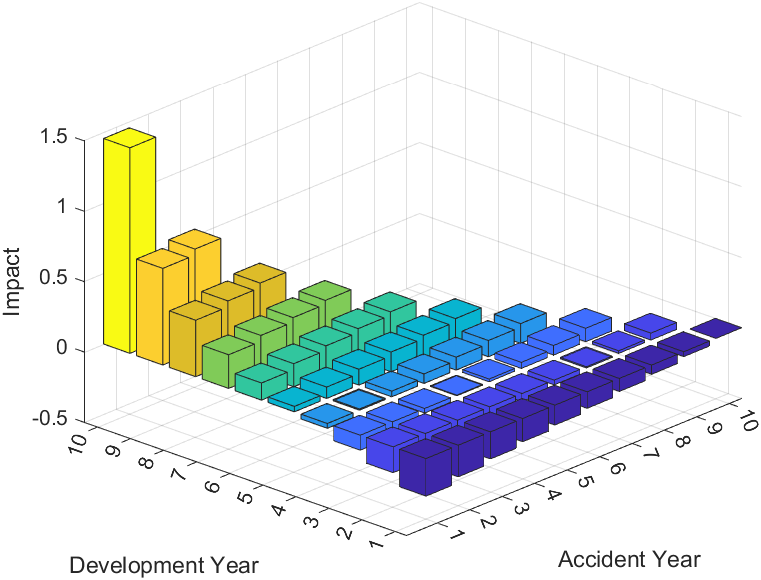}
			\caption{Illustration of $\text{IF}_{k,j}(\widehat{R}^{BF})$}
			\label{IFRBF}
		\end{center} 
			\end{figure}

	The impact function for total Bornhuetter-Ferguson reserves is given by $
\text{IF}_{k,j}(\widehat{R}^{BF})=\sum_{i=1}^{I}\text{IF}_{k,j}(\widehat{R}_i^{BF})$. The 3D graph for this impact function under the assumption that $\widehat{\mu}_i=\widehat{C}_{i,I}$ for all $i$ is given in Figure \ref{IFRBF}.
		Under this technique, the significant impact of $X_{I,1}$ is completely eliminated. Additionally, all impacts have been reduced in comparison to Figure \ref{VeVaDh091IFR} and in particular, the impact for late accident years at early development periods is significantly reduced. %\BAc{really?}
		This is because the denominator in equation \eqref{BFimp} for $k<i$ is greater for earlier development periods and the impact of $X_{k,j}$ on $\widehat{R}_i$ is zero for a greater proportion of accident years $i$ as we increase $k$. 
%	We now consider the impact functions for the mean squared error of reserves. 

\section{Impact on Mean Squared Error under Mack's Model}  \label{S_MSE}
\subsection{Individual Accident Year MSE} 
\begin{figure}[htb]
			\begin{center}
				\includegraphics[width=0.6\linewidth]{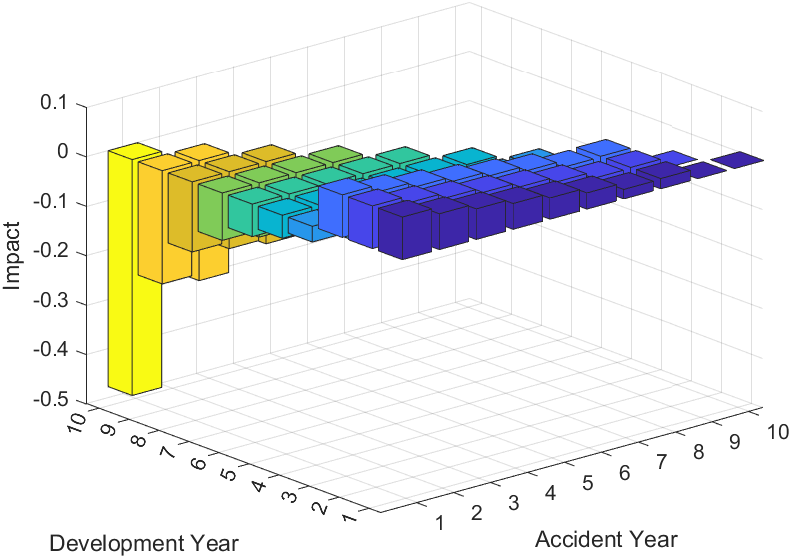}
				\caption{Illustration of $\text{IF}_{k,j}\left(\sqrt{\text{mse}(\widehat{R}_8)}\right)$}
				\label{IFmseR8}
			\end{center} 
		\end{figure}

	When calculating the impact function for this statistic we have considered the $\sigma_j$ and $f_j$ terms as known constants such that we are calculating the sensitivity of the mean squared error to incremental claims rather than the sensitivity of the estimate of this term. To approximate this function we may then then plug in the estimated values of $\sigma_j$ and $f_s$. The impact function is given by 
	\begin{eqnarray}\label{ifmse}
	\text{IF}_{k,j}(\text{mse}(\widehat{R}_i))=\begin{cases}
	0 &\text{if } k>i\\ \\
	\sum_{j=I-i+1}^{I-1}(f_{I-i+1}\cdot...\cdot f_{j-1}\sigma^2_jf_{j+1}^2\cdot...\cdot f_{I-1}^2)+\\2C_{i,I-i+1}(\widehat{f}_{I-i+1}\cdot...\cdot \widehat{f}_{I-1})^2\sum_{s=I-i+1}^{I-1}\frac{\widehat{\sigma}^2_s/\widehat{f}_s^2}{\sum_{i=1}^{I-s}C_{i,s}} & \text{if } k=i\\ \\
	-2C_{i,I-i+1}(\widehat{f}_{I-i+1}\cdot...\cdot \widehat{f}_{I-1})\sqrt{\sum_{s=I-i+1}^{I-1}\frac{\widehat{\sigma}^2_s/\widehat{f}_s^2}{\sum_{i=1}^{I-s}C_{i,s}}}\\\widehat{C}_{i,I}\sum_{p=k}^{i-1}\left(\left(\frac{1}{\sum_{i=1}^{p}C_{i,I-p+1}}\right)\textbf{1}_{\{j\leq I-p+1\}}-\left(\frac{1}{\sum_{i=1}^{p}C_{i,I-p}}\right)\textbf{1}_{\{j\leq I-p\}}\right) & \text{if } k\leq i-1 
	\end{cases}
	\end{eqnarray}
	See Appendix \ref{mseriproof} for proof. Note that as the results for $\text{IF}_{k,j}(\text{mse}(\widehat{R}_i))$ will be in units of \$$^2$ it is often desirable to look at the impact function for the root mean squared error (rmse). This is simply given by $
	\text{IF}_{k,j}\left(\sqrt{\text{mse}(\widehat{R}_i)}\right)=\frac{1}{2}\cdot\frac{\text{IF}_{k,j}(\text{mse}(\widehat{R}_i))}{\sqrt{\text{mse}(\widehat{R}_i)}}
$
and allows for the results given to be in the same units as reserves. % and this is what has been done for the illustration in Figure \ref{IFmseR8}.
	The impact that each incremental claim is having on the rmse of the reserves for accident year 8 is given in Table \ref{egifmseri} with a graphical representation provided in Figure \ref{IFmseR8}. From Equation \eqref{ifmse} we have that for $k=i$, the impact is always the sum of two positive terms and is independent of $j$. Hence for $k=i$ the impact is always positive and equal. This is represented by the row of equal height positive columns for accident year 8 in Figure \ref{IFmseR8}. Additionally, we observe that the sign of the impact is the opposite of that for $\text{IF}_{k,j}(\widehat{R}_8)$ (except when $k=8$) and we observe similar trends in terms of magnitude. In particular, note that the impact is increasing in magnitude towards the top right corner observation however these impacts are negative. 
		\begin{table}[htb]
		\begin{center}
			\resizebox{\linewidth}{!}{%
				\begin{tabular}{|c| c c c c c c c c c c|} \hline
					\textbf{$i$/$j$} & \textbf{1} & \textbf{2} & \textbf{3} &  \textbf{4} & \textbf{5} &  \textbf{6} & \textbf{7} & \textbf{8} & \textbf{9} & \textbf{10} \\ \hline
					\textbf{1} & 0.0863 & 0.0863 & 0.0863 & -0.0318 & -0.0468 & -0.0659 & -0.0960 & -0.1419 & -0.2291 & -0.4773\\ 
					\textbf{2} & 0.0724 & 0.0724 & 0.0724 & -0.0456 & -0.0606 & -0.0797 & -0.1099 & -0.1558 & -0.2429 &		   \\ 
					\textbf{3} & 0.0618 & 0.0618 & 0.0618 & -0.0562 & -0.0712 & -0.0903 & -0.1205 & -0.1664 &  &\\ 
					\textbf{4} & 0.0522 & 0.0522 & 0.0522 & -0.0658 & -0.0808 & -0.0999 & -0.1300 &  &  &	   \\ 
					\textbf{5} & 0.0430 & 0.0430 & 0.0430 & -0.0751 & -0.0900 & -0.1092 &  &  &  &   \\  
					\textbf{6} & 0.0327 & 0.0327 & 0.0327 & -0.0854 & -0.1004 &  &  &  &  &  \\ 
					\textbf{7} & 0.0193 & 0.0193 & 0.0193 & -0.0988 &  &  &  &  &  &  \\ 
					\textbf{8} & 0.0208 & 0.0208 & 0.0208 &  &  &  &  &  &  &  \\ 
					\textbf{9} & 0.0000 & 0.0000 &  &  &  &  &  &  &  &  \\ 
					\textbf{10} & 0.0000 &  &  &  &  &  &  &  &  &   \\ \hline
				\end{tabular}}
			\end{center}
			\caption{$\text{IF}_{k,j}\left(\sqrt{\text{mse}(\widehat{R}_8)}\right)$} 
			\label{egifmseri}
		\end{table}
		 For the cases when $k\leq i-1$ note that the term $-2C_{i,I-i+1}(\widehat{f}_{I-i+1}\cdot...\cdot \widehat{f}_{I-1})\sqrt{\sum_{s=I-i+1}^{I-1}\frac{\widehat{\sigma}^2_s/\widehat{f}_s^2}{\sum_{i=1}^{I-s}C_{i,s}}}$ is always negative and is independent of $k$ and $j$ (i.e. the same value for this term is used throughout the triangle for all $k\leq i-1$). Additionally, the term \\$\widehat{C}_{i,I}\sum_{\{p\in\mathbb{D}|p\geq k\}}\left(\left(\frac{1}{\sum_{i=1}^{p}C_{i,I-p+1}}\right)\textbf{1}_{\{j\leq I-p+1\}}-\left(\frac{1}{\sum_{i=1}^{p}C_{i,I-p}}\right)\textbf{1}_{\{j\leq I-p\}}\right)$ is equal to $\text{IF}_{k,j}(\widehat{R}_i)$ provided above. Hence we see that $\text{IF}_{k,j}(\text{mse}(\widehat{R}_i))$ will have opposite sign to $\text{IF}_{k,j}(\widehat{R}_i)$ throughout the triangle for $k\leq i-1$.\\\\Notably, for $k\leq i-1$ and $j\leq I-i+1$ the impact will be positive which is shown in Figure \ref{IFmseR8} for $k\leq 7$ and $j\leq 3$. We will see a change of sign for $\text{IF}_{k,j}(\text{mse}(\widehat{R}_i))$ from positive for $j\leq I-i+1$ to negative for $j=I-i+2$ when $k=i-1$. A proof of this property is given in Appendix \ref{signchange}.
Similar trends in terms of the magnitude are seen for $\text{IF}_{k,j}(\text{mse}(\widehat{R}_i))$ as was outlined above for $\text{IF}_{k,j}(\widehat{R}_i)$. For instance as we move towards the top right corner of the loss triangle we will tend to see the impact become increasingly negative (as opposed to increasingly positive for $\text{IF}_{k,j}(\widehat{R}_i)$). Further investigation into the relationship between $\text{IF}_{k,j}(\widehat{R}_i)$) and $\text{IF}_{k,j}(\text{mse}(\widehat{R}_i))$, particularly the change in sign when $k\leq i-1$ is a warranted extension of this work. 
\subsection{Estimate of total MSE}
\begin{figure}[htb]
		\begin{center}
			\includegraphics[width=0.6\linewidth]{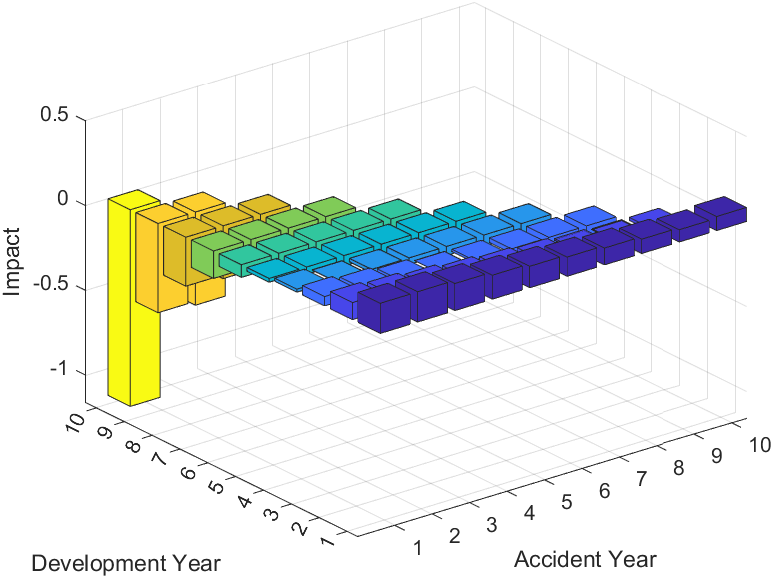}
			\caption{Illustration of $\text{IF}_{k,j}\left(\sqrt{\text{mse}(\widehat{R})}\right)$}
			\label{VeVaDh091IFmseR}
		\end{center} 
	\end{figure}

We have that for Mack's Model the mean squared error of prediction for total reserves is given by 
\begin{equation}
\text{mse}(\widehat{R})=\sum_{i=2}^{I}\left\lbrace(\text{s.e.}(\widehat{R}_i))^2+\widehat{C}_{i,I}\left(\sum_{q=i+1}^{I}\widehat{C}_{q,I}\right)\left(\sum_{r=I-i+1}^{I-1}\frac{2\sigma_r^2/\widehat{f}_r^2}{\sum_{n=1}^{I-r}C_{n,r}}\right)\right\rbrace 
\end{equation}
In this case, we are again considering the unknown $\sigma_r$ values as constants rather than taking their estimates. This again allows us to focus on the impact that incremental claims are having on the mean squared error rather than its associated estimate. 
The impact function is given by 
	\begin{eqnarray}
	\begin{split}
	\text{IF}_{k,j}\left(\widehat{\text{mse}(\widehat{R})}\right)=\sum_{i=2}^{I}\left\lbrace\widehat{\text{IF}}_{k,j}(\text{mse}(\widehat{R}_i))+\widehat{C}_{i,I}\left(\sum_{q=i+1}^{I}\widehat{C}_{q,I}\right)\sum_{r=I-i+1}^{I-1}\frac{-2\sigma_r^2\sum_{n=1}^{I-r}\widehat{f}_r^2C_{n,r}\left(\frac{\partial\ln  C_{n,r}}{\partial X_{k,j}}+\frac{2\partial\ln  \widehat{f}_r}{\partial X_{k,j}}\right)}{\left(\sum_{n=1}^{I-r}C_{n,r}\widehat{f}_r^2\right)^2}\right.+\\\left. \left(\sum_{r=I-i+1}^{I-1}\frac{2\sigma_r^2/\widehat{f}_r^2}{\sum_{n=1}^{I-r}C_{n,r}}\right)\left(\widehat{C}_{i,I}\left(\sum_{q=i+1}^{I}\left(\text{IF}_{k,j}(\widehat{R}_q)+\frac{\partial C_{q,I-q+1}}{\partial X_{k,j}}\right)\right)+\left(\sum_{q=i+1}^{I}\widehat{C}_{q,I}\right)\left(\text{IF}_{k,j}(\widehat{R}_i)+\frac{\partial C_{i,I-i+1}}{\partial X_{k,j}}\right)\right)\right\rbrace
	\end{split}
	\end{eqnarray}
Note that this formulation of the impact function still contains derivative terms which are readily calculable. 
Importantly, these impacts are not simply the sum of the impacts for the mse of each individual accident year. A similar point is made regarding taking the impact of the rmse for total reserves ($\sqrt{\text{mse}(\widehat{R})}$) when calculating impact functions in practice such that we are looking at the impact in the same units as reserves.
The impact of individual claims on the estimated rmse of total reserves is given in Figure \ref{VeVaDh091IFmseR}. It appears that the main result for these impacts is that for development period 1, all impacts are positive and then holding $k$ constant the impacts are decreasing with $j$ towards zero and then continuing in this pattern, are becoming increasingly negative towards the upper right corner. 
	 %We now provide the impact that incremental claims are having on the quantiles of total reserves based on an assumption that they follow a lognormal distribution.  

\newpage 

\section{Impact on Lognormal Quantiles} \label{S_quantiles}
\begin{figure}[htb]
			\begin{center}
				\includegraphics[width=0.6\linewidth]{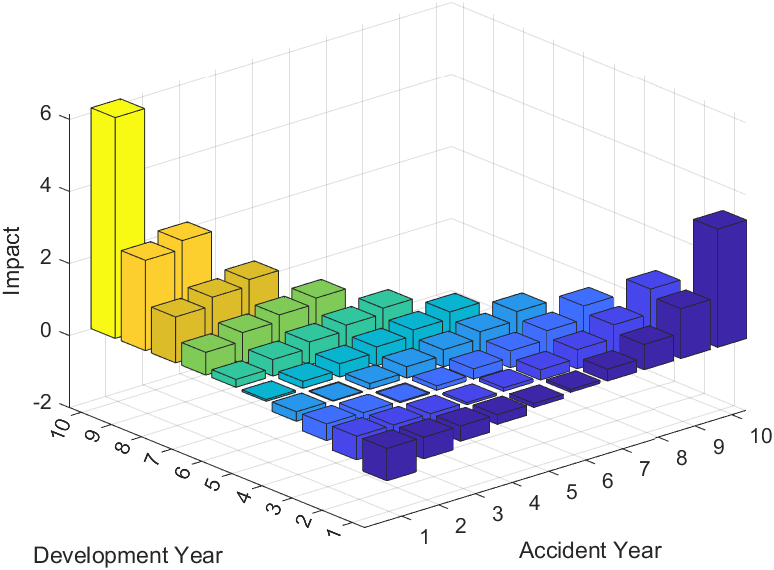}
				\caption{Illustration of $\text{IF}_{k,j}\left(F_R^{-1}(0.995)\right)$}
				\label{VeVaDh091IF995quantR}
			\end{center} 
		\end{figure}

	We now provide the impact function for total reserves under the common assumption that they are lognormally distributed. In principle, a similar approach may be employed for any location-scale distribution, though the mathematics of quantiles can become intractable here in the case of discrete distributions. For this reason, the over-dispersed Poisson distribution has been avoided, despite it being a more natural distribution to associate with the chain-ladder. An example given in Chapter 11 of \citep*{Tay00} shows (for that case) that quantiles other than extreme ones are little affected by the lognormal choice rather than a shorter-tailed distribution. Nonetheless,we would advise validation that lognormal is an appropriate choice for the data at hand before implementing the results provided here. 
	We have the following assumptions
	 \begin{equation} E[R]=\widehat{R}=e^{\mu+\frac{1}{2}\sigma^2} \quad \text{and} \quad  \text{Var}(R)=\text{mse}(\widehat{R})=e^{2\mu+\sigma^2}(e^{\sigma^2}-1)
	\end{equation}
	Such that $
	R\sim LN(\mu,\sigma^2)$. 
	The q quantile of a lognormal distribution, $X\sim LN(\mu,\sigma)$ is given by \begin{equation}
	F_X^{-1}(q)=e^{\mu+\sigma\Phi^{-1}(q)}
	\end{equation}
	Where $\Phi(.)$ is the cumulative distribution function of the standard normal distribution. The impact function for lognormal quantiles is given by 
	\begin{eqnarray}
	\begin{split}
	\text{IF}_{k,j}\left(F_R^{-1}(q)\right)=\left(\frac{2\cdot \text{IF}_{k,j}(\widehat{R})\cdot\widehat{R}-\text{IF}_{k,j}\left(\text{mse}(\widehat{R})\right)}{2(\text{mse}(\widehat{R})+\widehat{R}^2)}+\frac{\Phi^{-1}(q)\left(\text{IF}_{k,j}(\text{mse}(\widehat{R}))\cdot\widehat{R}-2\text{mse}(\widehat{R})\cdot \text{IF}_{k,j}(\widehat{R})\right)}{2\widehat{R}\left(\text{mse}(\widehat{R})+\widehat{R}^2\right)\sqrt{\ln \left(1+\frac{\text{mse}\widehat{R}}{\widehat{R}^2}\right)}}\right)\times \\
	\exp\left[\ln (\widehat{R})-\frac{1}{2}\ln \left(1+\frac{\text{mse}(\widehat{R})}{\widehat{{R}^2}}\right)+\sqrt{\ln \left(1+\frac{\text{mse}(\widehat{R})}{\widehat{{R}^2}}\right)}\Phi^{-1}(q)\right]
	\end{split}
	\end{eqnarray}
See Appendix \ref{quantilesproof} for proof. 	 The impact that each incremental claim is having on the 99.5\% quantile of total reserves under the assumption that they are log-normally distributed is given in Table \ref{egifquant} and the corresponding 3D graph is given in Figure \ref{VeVaDh091IF995quantR}. 
Importantly, we see similar trends in this impact triangle as were seen for $\text{IF}_{k,j}(\widehat{R})$ (Figure \ref{VeVaDh091IFR}). Notably, the three cornerpoints $X_{1,1}$, $X_{1,10}$ and $X_{10,1}$ are having significant impacts on the 99.5\% quantile of reserves. This can be understood intuitively in that if an incremental claim is having a given impact on reserves then we may expect to see a similar impact on their associated quantiles. % We now describe the robust bivariate chain-ladder technique and illustrate two alternative formulations of this methodology. %Interestingly, in the examples we have considered the features of the impact function for quantiles are closely related to what we have observed for the impact function of total reserves ($\text{IF}_{k,j}(\widehat{R})$) (given $q>0.5$) as was illustrated in Figure \ref{IFR}. 		
		\begin{table}[htb]
		\begin{center}
			\resizebox{\linewidth}{!}{%
				\begin{tabular}{|c| c c c c c c c c c c|} \hline
					\textbf{$i$/$j$} & \textbf{1} & \textbf{2} & \textbf{3} &  \textbf{4} & \textbf{5} &  \textbf{6} & \textbf{7} & \textbf{8} & \textbf{9} & \textbf{10} \\ \hline
					\textbf{1} & -0.8366 & -0.6309 & -0.4455 & -0.2543 & -0.0435 & 0.2220 & 0.6332 & 1.2749 & 2.5060 & 6.1290	\\ 
					\textbf{2} & -0.6009 & -0.3952 & -0.2098 & -0.0186 & 0.1922 & 0.4576 & 0.8689 & 1.5106 & 2.7417 & \\ 
					\textbf{3} & -0.4235 & -0.2179 & -0.0325 & 0.1587 & 0.3695 & 0.6350 & 1.0462 & 1.6880 &  & \\ 
					\textbf{4} & -0.2677 & -0.0620 & 0.1234 & 0.3146 & 0.5254 & 0.7908 & 1.2021 &  &  &  \\ 
					\textbf{5} & -0.1088 & 0.0968 & 0.2823 & 0.4734 & 0.6842 & 0.9497 &  &  &  &  \\  
					\textbf{6} & 0.0786 & 0.2842 & 0.4696 & 0.6608 & 0.8716 &  &  &  &  & \\ 
					\textbf{7} & 0.3274 & 0.5330 & 0.7185 & 0.9096 &  &  &  &  &  & \\ 
					\textbf{8} & 0.7020 & 0.9077 & 1.0931 &  &  &  &  &  &  & \\ 
					\textbf{9} & 1.3842 & 1.5898 &  &  &  &  &  &  &  & \\ 
					\textbf{10} & 3.2803 &  &  &  &  &  &  &  &  &  \\ \hline
				\end{tabular}}
			\end{center}
			\caption{$\text{IF}_{k,j}\left(F_R^{-1}(0.995)\right)$} 
			\label{egifquant}
		\end{table}

\clearpage		
		
\section{Conclusion}\label{conclusion}

In this paper we have provided impact functions for a range of statistics of interest under Mack's Model as well as reserves under the Bornhuetter-Ferguson technique. Properties of these impact functions have been discussed and we have illustrated their calculation on real data. These impact functions capture the rate at which the relevant statistic of interest will change given movement in a particular incremental claim. Additionally, we can highlight the marginal contribution of each incremental claim to reserves as they are homogeneous functions of order one. 

We have illustrated that there is often a small set of observations that these statistics are particularly sensitive to. This highlights a lack of robustness in that deviations in some observations may largely dictate results. A further feature of these functions is that they are heavily dependent on the numerous other observations within a loss triangle, highlighting the interdependence of the claims development and each incremental observation. Additionally, all impact functions that have been derived in this section are unbounded with respect to individual incremental claims except for the cases when they equal zero. Hence the relevant statistics of interest may be carried arbitrarily far from their true value in the presence of outlying observations.

Finally, we have illustrated results using data from a Belgian non-life insurer.

\section*{Code}

The \verb_R_ code to replicate numerical results and figures is available on the GitHub repository \url{https://github.com/agi-lab/reserving-impact-factors}.

\section*{Acknowledgements}
Earlier versions of this paper were presented at the Australian Actuaries Institute \emph{General Insurance Seminar} conference in Melbourne (Australia), as well as at an Australian  Actuaries Institute \emph{Insights} session in Sydney (Australia). The authors are grateful for constructive comments received from colleagues who attended this event and from those who read the earlier version of the paper published online to support the presentation. The authors are also grateful to Sharanjit Paddam and Will Turvey for comments on earlier drafts of the paper, as well as William Cheng, Wanzhang (Simon) Jing, and Yun Wai (William) Ho for their research assistance.

This research was supported under Australian Research Council's Linkage (LP130100723, with funding partners Allianz Australia Insurance Ltd, Insurance Australia Group Ltd, and Suncorp Metway Ltd) and Discovery (DP200101859) Projects funding schemes. Furthermore, Lavender acknowledges support from the UNSW Business School Honours Scholarship. The views expressed herein are those of the authors and are not necessarily those of the supporting organisations.  

\section*{References}

\bibliographystyle{elsarticle-harv}
\bibliography{libraries}

\appendix

\section*{Appendix}
\section{Data}\label{data}\label{unidata}
%\subsection{Univariate} \label{unidata}
To illustrate the results of our impact functions we use data from a Belgian non-life insurer as given in \citet*{VeVaDh09} and presented in Table \ref{VeVaDh091} in incremental form. This triangle exhibits a long-tail and stable results over subsequent accident years. Results for triangles with differing characteristics, (for example, for a developing portfolio where the difference in volumes of consequent accident years can be significant) are expected to lead to different observed impacts, notwithstanding the universal properties of impact functions across all triangles outlined in this paper.  
\begin{table}[htb]
	\begin{center}
		\resizebox{1\linewidth}{!}{%
			\begin{tabular}{|c|cccccccccc|} \hline
				\textbf{$i$/$j$} & \textbf{1} & \textbf{2} & \textbf{3} &  \textbf{4} & \textbf{5} &  \textbf{6} & \textbf{7} & \textbf{8} & \textbf{9} & \textbf{10}\\ \hline
				\textbf{1} &135 338 126&	90 806 681&	68 666 715&	55 736 215&	46 967 279&	35 463 367&	30 477 244&	24 838 121&	18 238 489&	14 695 083 \\ 
				\textbf{2} &125 222 434&	89 639 978&	70 697 962&	58 649 114&	46 314 227&	41 369 299&	34 394 512&	26 554 172&	24 602 209 &    \\ 
				\textbf{3}&136 001 521&	91 672 958&	78 246 269&	62 305 193&	49 115 673&	36 631 598&	30 210 729&	29 882 359 & & \\ 
				\textbf{4} &135 277 744&	103 604 885&	78 303 084&	61 812 683&	48 720 135&	39 271 861&	32 029 697& & &   \\ 
				\textbf{5} &143 540 778&	109 316 613&	79 092 473&	65 603 900&	51 226 270&	44 408 236 & & & &   \\  
				\textbf{6} &132 095 863&	88 862 933&	69 269 383&	57 109 637&	48 818 781& & & & &   \\ 
				\textbf{7} &127 299 710&	92 979 311&	61 379 607	&50 317 305 &  & & & & & \\ 
				\textbf{8} &120 660 241&	89 469 673&	71 570 718 &  &  & & & & &\\ 
				\textbf{9} &134 132 283&	87 016 365 &  &  &  & & & & &  \\ 
				\textbf{10} &131 918 566  &  &  &  &  & & & & &  \\ \hline
			\end{tabular}}
		\end{center}
		\caption{Incremental Claims Data from \cite{VeVaDh09}} 
		\label{VeVaDh091}
	\end{table}

\section{Proofs}\label{S:proof}
In this chapter we provide proofs for the impact functions and their properties that were presented in Chapter \ref{impacts}. Note that all proofs are based on an assumption of non-negative incremental claims $X_{k,j}$. 
\subsection{Accident Year Reserves $\text{IF}_{k,j}(\widehat{R}_i)$}\label{ifriproof}
We have that \begin{equation}
	\widehat{R}_i=C_{i,I-i+1}(\widehat{f}_{I-i+1}\cdots\widehat{f}_{I-1})-C_{i,I-i+1}
\end{equation}
Write \begin{eqnarray*}
	f(X)	& =	& C_{i,I-i+1}(\widehat{f}_{I-i+1}\cdots\widehat{f}_{I-1}) \\ 
	\ln(f(X))	& =	&\ln(C_{i,I-i+1})+\ln(\widehat{f}_{I-i+1})+\cdots +\ln(\widehat{f}_{I-1})\\
	\frac{1}{f(X)}\frac{\partial f(X)}{\partial X_{k,j}}	& =	&	\frac{\partial \ln(C_{i,I-i+1})}{\partial X_{k,j}}+\frac{\partial \ln(\widehat{f}_{I-i+1})}{\partial X_{k,j}}+\cdots +\frac{\partial \ln(\widehat{f}_{I-1})}{\partial X_{k,j}} \\
	\frac{\partial f(X)}{\partial X_{k,j}}	& =	& f(X)\left(	\frac{\partial \ln(C_{i,I-i+1})}{\partial X_{k,j}}+\frac{\partial \ln(\widehat{f}_{I-i+1})}{\partial X_{k,j}}+\cdots +\frac{\partial \ln(\widehat{f}_{I-1})}{\partial X_{k,j}}\right) \\
	& \therefore	&
\end{eqnarray*}
\begin{equation}\label{ifri}
	\frac{\partial \widehat{R}_i}{\partial X_{k,j}}=\widehat{C}_{i,I}\left(\frac{\partial \ln(C_{i,I-i+1})}{\partial X_{k,j}}+\frac{\partial \ln(\widehat{f}_{I-i+1})}{\partial X_{k,j}}+\cdots+\frac{\partial \ln(\widehat{f}_{I-1})}{\partial X_{k,j}}\right)-\frac{\partial}{\partial X_{k,j}}C_{i,I-i+1}
\end{equation}
We have that 
\begin{eqnarray}\label{ifc}
	\frac{\partial}{\partial X_{k,j}}C_{i,I-i+1}=\begin{cases}
		1, &\text{if }k=i \\
		0, &\text{otherwise}
	\end{cases}
\end{eqnarray}
\begin{eqnarray}
	\frac{\partial \ln(C_{i,I-i+1})}{\partial X_{k,j}}
	& =	& \frac{\partial}{\partial X_{k,j}}\ln(X_{i,1}+\cdots+X_{i,I-i+1}) = \begin{cases}\label{iflnc}
		\frac{1}{C_{i,I-i+1}}, &\text{if } k=i \\
		0, &\text{otherwise}
	\end{cases}
\end{eqnarray}
Now for $s\in{I-i+1,\dots,I-1}$, 
\begin{eqnarray}
	\frac{\partial}{\partial X_{k,j}}\ln(\widehat{f}_s)	& =	&\frac{\partial}{\partial X_{k,j}}\ln\left(\frac{\sum_{i=1}^{I-s}C_{i,s+1}}{\sum_{i=1}^{I-s}C_{i,s}}\right)\\
	& =	& \frac{\partial}{\partial X_{k,j}}(\ln\left(X_{1,1}+\cdots+X_{1,s+1}+\cdots+X_{I-s,1}+\cdots X_{I-s,s+1})\right. \label{iffs}\\ 
	& -	& \left.\ln(X_{1,1}+\cdots+X_{1,s}+\cdots+X_{I-s,1}+\cdots X_{I-s,s})\right) \nonumber
\end{eqnarray}
Note that equation \eqref{iffs} is equal to zero if $k>I-s$. The largest value $I-s$ can take is $I-I+i-1=i-1$. The first two cases of the impact function (equation \eqref{IFRi}) follow from equations \eqref{ifc}, \eqref{iflnc} and \eqref{iffs}). \\[1ex] We now derive the result for $k \leq i-1$. In these cases the impact function is given by 
\begin{eqnarray}
	\widehat{C}_{i,I}\cdot\frac{\partial}{\partial X_{k,j}}\left(\sum_{s=I-i+1}^{I-1}\ln(\widehat{f}_s)\right)
\end{eqnarray}
First consider when $s=I-i+1$. 
\begin{eqnarray}\label{iffhatj}
	\frac{\partial}{\partial X_{k,j}}\widehat{f}_{I-i+1}	& =	& \frac{\partial}{\partial X_{k,j}}\left(\ln(X_{1,1}+\cdots+X_{1,I-i+2}+\right.\\ \left.\cdots+X_{i-1,1}+\cdots X_{i-1,I-i+2}\right)
	& -	& \ln(X_{1,1}+\cdots+X_{1,I-i+1}+\cdots+X_{i-1,1}+\cdots X_{i-1,I-i+1})) \nonumber
\end{eqnarray}
Notice that $X_{k,j}$ will appear once only in the first logarithmic term on the right hand side of equation \eqref{iffhatj} if $j=I-i+2$. However it will appear once in both terms if $j\leq I-i+1$. Hence we have the following form for the cases when $s=I-i+1$.
\begin{eqnarray}\label{iflnfhatj}
	\frac{\partial}{\partial X_{k,j}}\ln(\widehat{f}_{I-i+1}) =
	\begin{cases}
		\frac{1}{\sum_{q=1}^{i-1}C_{q,I-i+2}}-\frac{1}{\sum_{q=1}^{i-1}C_{q,I-i+1}}, & \text{if } j\leq I-i+1 \\
		\frac{1}{\sum_{q=1}^{i-1}C_{q,I-i+2}}, & \text{if } j=I-i+2
	\end{cases}
\end{eqnarray}
A similar result can be obtained for $s=I-i+2$ such that 
\begin{eqnarray}\label{iflnfhatj2}
	\frac{\partial}{\partial X_{k,j}}\ln(\widehat{f}_{I-i+2})= \begin{cases} 
		\frac{1}{\sum_{q=1}^{i-2}C_{q,I-i+3}}-\frac{1}{\sum_{q=1}^{i-2}C_{q,I-i+2}}, &\text{if } j\leq I-i+2 \\
		\frac{1}{\sum_{q=1}^{i-2}C_{q,I-i+3}},& \text{if } j=I-i+3
	\end{cases}
\end{eqnarray}
An important point to note is that for each development factor, we are summing the cumulative payments in the denominator(s) up to $I-s$. In equation \eqref{iflnfhatj} this is to $i-1$ such that all values of $k\leq i-1$ are included. However in equation \eqref{iflnfhatj2} this is only up to $i-2$. This implies that if $k=i-1$ then $X_{k,j}$ will only appear in the first development factor in equation \eqref{ifri}. This pattern will continue such that for $k=i-2$, $X_{k,j}$ will only appear in the first two developments factors and so on. For $k=1$, $X_{k,j}$ will appear in all development factors. This completes the proof. 
\subsection{Properties of $\text{IF}_{k,j}(\widehat{R}_i)$}\label{ifriprops} 
\begin{enumerate} \item$\text{IF}_{k,j}(\widehat{R}_i)\leq0$ for all $k\leq i-1$ and $j\leq I-i+1$. We have \begin{equation}\label{indicator}
		\text{IF}_{k,j}(\widehat{R}_i)=\widehat{C}_{i,I}\sum_{p=k}^{i-1}\left(\left(\frac{1}{\sum_{q=1}^{p}C_{q,I-p+1}}\right)\textbf{1}_{\{j\leq I-p+1\}}-\left(\frac{1}{\sum_{q=1}^{p}C_{q,I-p}}\right)\textbf{1}_{\{j\leq I-p\}}\right)
	\end{equation} 
	Where empty sums equal zero. Note that the smallest value $I-p$ can take is $I-i+1$. Hence for $j\leq I-i+1$ the indicator functions in equation \eqref{indicator} are always equal to 1. Now we can turn to an arbitrary case of the term inside the summand (i.e. for any p). We have \begin{eqnarray}
		\frac{1}{\sum_{q=1}^{p}C_{q,I-p+1}}-\frac{1}{\sum_{q=1}^{p}C_{q,I-p}} 
		& =	&\frac{1}{\sum_{q=1}^{p}C_{q,I-p}+\sum_{q=1}^{p}X_{q,I-p+1}}-\frac{1}{\sum_{q=1}^{p}C_{q,I-p}}\\
		& \leq	& 0
	\end{eqnarray}
	Hence $\text{IF}_{k,j}(\widehat{R}_i)$ will be a sum of non-positive terms when $k\leq i-1$ and $j\leq I-i+1$ and hence $\text{IF}_{k,j}(\widehat{R}_i)\leq0$. \item From property 1 above we know that $\frac{1}{\sum_{q=1}^{p}C_{q,I-p+1}}-\frac{1}{\sum_{q=1}^{p}C_{q,I-p}}\leq 0$. Now for $j>I-i+1$ we have \begin{equation}
		\text{IF}_{k,j}(\widehat{R}_i)=\widehat{C}_{i,I}\left(\sum_{\{p\in\mathbb{D}|p\geq k\ \cap\ p<I-j+1\}}\left(\left(\frac{1}{\sum_{q=1}^{p}C_{q,I-p+1}}\right)-\left(\frac{1}{\sum_{q=1}^{p}C_{q,I-p}}\right)\right)+\left(\frac{1}{\sum_{q=1}^{I-j+1}C_{q,j}}\right)\right)
	\end{equation}
	Where $\mathbb{D}=\{1,...,i-1\}$. Hence for these cases $\text{IF}_{k,j}(\widehat{R}_i)>0$ if\\ $-\sum_{\{p\in\mathbb{D}|p\geq k\ \cap\ p<I-j+1\}}\left(\left(\frac{1}{\sum_{q=1}^{p}C_{q,I-p+1}}\right)-\left(\frac{1}{\sum_{q=1}^{p}C_{q,I-p}}\right)\right)<\left(\frac{1}{\sum_{q=1}^{I-j+1}C_{q,j}}\right)$.
	\item For fixed $j$, $\text{IF}_{k,j}(\widehat{R}_i)$ is non-decreasing with $k$. For $k=i$ we know that the impact is always non-negative and from Property 1 we know that $\text{IF}_{k,j}\leq0$ for all $k\leq i-1$ and $j\leq I-i+1$ and hence for fixed $j$ $\text{IF}_{i,j}(\widehat{R}_i)\geq \text{IF}_{i-1,j}(\widehat{R}_i)$. For $k\leq i-1$ we have that \begin{equation}
		\text{IF}_{k,j}(\widehat{R}_i)=\widehat{C}_{i,I}\sum_{p=k}^{i-1}\left(\left(\frac{1}{\sum_{q=1}^{p}C_{q,I-p+1}}\right)\textbf{1}_{\{j\leq I-p+1\}}-\left(\frac{1}{\sum_{q=1}^{p}C_{q,I-p}}\right)\textbf{1}_{\{j\leq I-p\}}\right)
	\end{equation} 
	Now by fixing $j$ note that the indicator functions will take the same value independent of $k$. However as we increase $k$ then the number of terms we are summing is decreased by the condition that $p\geq k$. Now for $j\leq I-i+1$, from Property 1 we know that $\text{IF}_{k,j}(\widehat{R}_i)$ is made up of the sum of non-positive terms. Hence as we increase with $k$ the number of negative terms we are summing over decreases and hence we see $\text{IF}_{k,j}(\widehat{R}_i)$ increase. Similarly, for $j>I-i+1$, from Property 2 we have\\ $
	\text{IF}_{k,j}(\widehat{R}_i)=\widehat{C}_{i,I}\left(\sum_{\{p\in\mathbb{D}|p\geq k\ \cap\ p<I-j+1\}}\left(\left(\frac{1}{\sum_{q=1}^{p}C_{q,I-p+1}}\right)-\left(\frac{1}{\sum_{q=1}^{p}C_{q,I-p}}\right)\right)+\left(\frac{1}{\sum_{q=1}^{I-j+1}C_{q,j}}\right)\right)
	$.
	Again as we increase k, the number of negative terms we are summing before adding the positive term (which remains the same irrespective of $k$) is reduced and this is why we see $\text{IF}_{k,j}(\widehat{R}_i)$ increase with $k$ for fixed $j$.
	\item	We consider cases for $j=I-k+1$ (i.e. the most recent diagonal). For $k=i-1$ we have \begin{equation}
		\text{IF}_{k,j}(\widehat{R}_i)=\frac{\widehat{C}_{i,I}}{\sum_{q=1}^{i-1}C_{q,I-i+2}}\quad \text{and  }\quad \text{IF}_{k+1,j-1}(\widehat{R}_i)=(\widehat{f}_{I-i+1}\cdot...\cdot\widehat{f}_{I-1}-1)
	\end{equation}
	Such that $\text{IF}_{k,j}(\widehat{R}_i)>\text{IF}_{k+1,j-1}(\widehat{R}_i)$  if  $\widehat{C}_{i,I}>(\widehat{f}_{I-i+1}\cdot...\cdot\widehat{f}_{I-1}-1)\cdot \sum_{q=1}^{i-1}C_{q,I-i+2}$. For $k\leq i-2$ we have \begin{equation}
		\text{IF}_{k,j}(\widehat{R}_i)=\widehat{C}_{i,I}\frac{1}{\sum_{q=1}^{k}C_{q,j}}
		\quad \text{and } \quad 
		\text{IF}_{k+1,j-1}(\widehat{R}_i)=\widehat{C}_{i,I}\frac{1}{\sum_{q=1}^{k+1}C_{q,j-1}}
	\end{equation}
	Hence for $\text{IF}_{k,j}(\widehat{R}_i)>\text{IF}_{k+1,j-1}(\widehat{R}_i)$ to hold we require that \begin{eqnarray}
		\frac{1}{\sum_{q=1}^{k}C_{q,j}}& >	&	\frac{1}{\sum_{q=1}^{k+1}C_{q,j-1}}\\
		\frac{1}{C_{1,j-1}+\cdots+C_{k,j-1}+X_{1,j}+\cdots+X_{k,j}}& >	&\frac{1}{C_{1,j-1}+\cdots +C_{k,j-1}+C_{k+1,j-1}}\\
		\frac{1}{C_{1,j-1}+\cdots+C_{k,j-1}+ \sum_{q=1}^{k}X_{q,j}}& >	&\frac{1}{C_{1,j-1}+\cdots +C_{k,j-1}+C_{k+1,j-1}}
	\end{eqnarray}
	From here we can see that this inequality will only be true if  
	$\sum_{q=1}^{k}X_{q,j}<C_{k+1,j-1}$.
	\item Unfortunately, for the diagonals that do not represent the most recent calendar year a similarly succinct result is not available. Consider $k\leq i-3$ and $j=I-k$ (i.e. the second most recent diagonal of the loss reserve triangle). In this case we have \begin{equation}\label{diag1}
		\text{IF}_{k,j}(\widehat{R}_i)=\widehat{C}_{i,I}\left(\frac{1}{\sum_{q=1}^{k}C_{q,j+1}}-\frac{1}{\sum_{q=1}^{k}C_{q,j}}+\frac{1}{\sum_{q=1}^{k+1}C_{q,j}}\right)
	\end{equation}
	and \begin{equation}\label{diag2}
		\text{IF}_{k+1,j-1}(\widehat{R}_i)=\widehat{C}_{i,I}\left(\frac{1}{\sum_{q=1}^{k+1}C_{q,j}}-\frac{1}{\sum_{q=1}^{k+1}C_{q,j-1}}+\frac{1}{\sum_{q=1}^{k+2}C_{q,j-1}}\right)
	\end{equation}
	Now for $\text{IF}_{k,j}(\widehat{R}_i)>\text{IF}_{k+1,j-1}(\widehat{R}_i)$ to hold we require 
	\begin{equation}
		\frac{1}{\sum_{q=1}^{k}C_{q,j+1}}-\frac{1}{\sum_{q=1}^{k}C_{q,j}}>\frac{1}{\sum_{q=1}^{k+2}C_{q,j-1}}-\frac{1}{\sum_{q=1}^{k+1}C_{q,j-1}}
	\end{equation}
	Focusing on the left hand side we have that \begin{equation}
		\frac{1}{\sum_{q=1}^{k}C_{q,j+1}}-\frac{1}{\sum_{q=1}^{k}C_{q,j}}=\frac{1}{\sum_{q=1}^{k}C_{q,j}+\sum_{q=1}^{k}X_{q,j+1}}-\frac{1}{\sum_{q=1}^{k}C_{q,j}}
	\end{equation}
	This term will be non-positive and hence will be maximised by minimising $\sum_{q=1}^{k}X_{q,j+1}$ such that it will equal zero if $\sum_{q=1}^{k}X_{q,j+1}=0$. Now we turn to the right hand side where 
	\begin{equation}
		\frac{1}{\sum_{q=1}^{k+2}C_{q,j-1}}-\frac{1}{\sum_{q=1}^{k+1}C_{q,j-1}}=\frac{1}{\sum_{q=1}^{k+1}C_{q,j-1}+C_{k+2,j-1}}-\frac{1}{\sum_{q=1}^{k+1}C_{q,j-1}}
	\end{equation}
	This term will also be non-positive and hence will be maximised by minimising $C_{k+2,j-1}$ such that it will equal zero if $C_{k+2,j-1}=0$. Hence it can be seen that the inequality $\text{IF}_{k,j}(\widehat{R}_i)>\text{IF}_{k+1,j-1}(\widehat{R}_i)$ will be violated if $C_{k+2,j-1}=0$ and $\sum_{q=1}^{k}X_{q,j+1}>0$. However these are not the only conditions at which the inequality may be violated and in fact it may be violated when $C_{k+2,j-1}>\sum_{q=1}^{k}X_{q,j+1}$. \\[1ex] As we continue to move backward diagonally similar results can be found. For $k\leq i-4$ and $j=I-k-1$ we have that \begin{equation}
		\text{IF}_{k,j}(\widehat{R}_i)=\widehat{C}_{i,I}\left(\frac{1}{\sum_{q=1}^{k}C_{q,j+2}}-\frac{1}{\sum_{q=1}^{k}C_{q,j+1}}+\frac{1}{\sum_{q=1}^{k+1}C_{q,j+1}}-\frac{1}{\sum_{q=1}^{k+1}C_{q,j}}+\frac{1}{\sum_{q=1}^{k+2}C_{q,j}}\right)
	\end{equation}
	And 
	\begin{equation}
		\text{IF}_{k+1,j-1}(\widehat{R}_i)=\widehat{C}_{i,I}(\frac{1}{\sum_{q=1}^{k+1}C_{q,j+1}}-\frac{1}{\sum_{q=1}^{k+1}C_{q,j}}+\frac{1}{\sum_{q=1}^{k+2}C_{q,j}}-\frac{1}{\sum_{q=1}^{k+2}C_{q,j-1}}+\frac{1}{\sum_{q=1}^{k+3}C_{q,j-1}})
	\end{equation}
	Such that for $\text{IF}_{k,j}(\widehat{R}_i)>\text{IF}_{k+1,j-1}(\widehat{R}_i)$ to hold we require \begin{equation}
		\frac{1}{\sum_{q=1}^{k}C_{q,j+2}}-\frac{1}{\sum_{q=1}^{k}C_{q,j+1}}>\frac{1}{\sum_{q=1}^{k+3}C_{q,j-1}}-\frac{1}{\sum_{q=1}^{k+2}C_{q,j-1}}
	\end{equation}
	Again we can see that the inequality will be violated if $C_{k+3,j-1}=0$ and $\sum_{q=1}^{k}X_{q,j+2}>0$. Once again this is not the only point that the inequality is violated. 
	\\[1ex]A generalised result for all diagonals excluding the most recent is that for $\text{IF}_{k,j}(\widehat{R}_i)>\text{IF}_{k+1,j-1}(\widehat{R}_i)$ to hold we require that  \begin{equation}
		\frac{1}{\sum_{q=1}^{k}C_{q,I-k+1}}-\frac{1}{\sum_{q=1}^{k}C_{q,I-k}}>\frac{1}{\sum_{q=1}^{k+l}C_{q,I-k+1-l}}-\frac{1}{\sum_{q=1}^{k+l-1}C_{q,I-k+1-l}}
	\end{equation}
	Where $l$ represents the diagonal that is being evaluated. 
	\item For fixed $k$, $\text{IF}_{k,j}(\widehat{R}_i)$ is increasing with $j$ for $j\geq I-i+1$. In this case we have that \begin{equation}
		\text{IF}_{k,j}(\widehat{R}_i)=\widehat{C}_{i,I}\sum_{p=k}^{i-1}\left(\left(\frac{1}{\sum_{q=1}^{p}C_{q,I-p+1}}\right)\textbf{1}_{\{j\leq I-p+1\}}-\left(\frac{1}{\sum_{q=1}^{p}C_{q,I-p}}\right)\textbf{1}_{\{j\leq I-p\}}\right)
	\end{equation}
	Note that for fixed $k$, we will be summing over the same $p$ values. However as we increase $j$, fewer terms will satisfy the conditions of the indicator functions. Firstly consider $j=I-i+1$ against $j=I-i+2$. In this case, the negative term inside the summand for $p=i-1$ will be there for $j=I-i+1$ however will be absent for $j=I-i+2$ and hence we will see an increase in the impact function. This situation continues as we increase $j$, however for $j>I-i+1$ two terms will be absent for each incremental increase of $j$ rather than only the single negative term when comparing $j=I-i+1$ against $j=I-i+2$. For instance, comparing $j=I-i+2$ against $j=I-i+3$, the negative term inside the summand corresponding to the case when $p=i-2$ will be absent for $j=I-i+3$. However the positive term inside the summand corresponding to $p=i-1$ will also be absent (i.e. the case for $p=i-1$ will be equal to zero). Notably the difference between the case when $j=I-i+2$ and $j=I-i+3$ will be the removal of $\frac{1}{\sum_{q=1}^{i-1}C_{q,I-i+2}}-\frac{1}{\sum_{q=1}^{i-2}C_{q,I-i+2}}\leq0$ This pattern will continue as we increase $j$. A general result is that \begin{equation}
		\text{IF}_{k,j+1}(\widehat{R}_i)-\text{IF}_{k,j}(\widehat{R}_i)=\widehat{C}_{i,I}\left(\frac{1}{\sum_{q=1}^{I-j}C_{q,j}}-\frac{1}{\sum_{q=1}^{I-j+1}C_{q,j}}\right)\geq0
	\end{equation} 
\end{enumerate}
\textbf{Remark 1. }From properties 5 and 6 we have that $\text{IF}_{k_1,j_1}(\widehat{R}_i)\geq \text{IF}_{k_2,j_2}(\widehat{R}_i)$ whenever $(k_1,j_1)\succ(k_2,j_2)$ where $\succ$ represents the partial ordering such that for $(k_1,j_1)\succ(k_2,j_2)$ we require that $k_1\geq k_2$ and $j_1\geq j_2$ and $(k_1,j_1)\neq(k_2,j_2)$. Note that the strict inequality $\text{IF}_{k_1,j_1}(\widehat{R}_i)> \text{IF}_{k_2,j_2}(\widehat{R}_i)$ will hold when $j_1,j_2\geq I-i+1$ or when $k_1>k_2$. 
\subsection{$\text{IF}_{k,j}^{BF}(\widehat{R}_i)$}\label{bfproof}
We have that \begin{eqnarray}
\text{IF}_{k,j}(\widehat{R}_i^{BF})=\frac{\partial}{\partial X_{k,j}}(\widehat{R}_i^{BF}) 
=\frac{\partial}{\partial X_{k,j}}\left(\widehat{\mu}_i-\widehat{\mu}_i\frac{1}{\widehat{f}_{I-i+1}\cdot...\cdot\widehat{f}_{I-1}}\right)=-\widehat{\mu}_i\frac{\partial}{\partial X_{k,j}}\left(\frac{1}{\widehat{f}_{I-i+1}\cdot...\cdot\widehat{f}_{I-1}}\right)
\end{eqnarray}
From the previous derivation of the impact function for the CL reserve estimates we have that \begin{equation}
\begin{split}
\frac{\partial}{\partial X_{k,j}}(\widehat{f}_{I-i+1}\cdot...\cdot\widehat{f}_{I-1})=(\widehat{f}_{I-i+1}\cdot...\cdot\widehat{f}_{I-1})\sum_{p=k}^{i-1}\left(\left(\frac{1}{\sum_{q=1}^{p}C_{q,I-p+1}}\right)\textbf{1}_{\{j\leq I-p+1\}}\right.-\\\left.\left(\frac{1}{\sum_{q=1}^{p}C_{q,I-p}}\right)\textbf{1}_{\{j\leq I-p\}}\right)
\end{split}
\end{equation}
Hence we have \begin{eqnarray}
-\widehat{\mu}_i\frac{\partial}{\partial X_{k,j}}\left(\frac{1}{\widehat{f}_{I-i+1}\cdot...\cdot\widehat{f}_{I-1}}\right)	& =	& -\widehat{\mu}_i\frac{\partial}{\partial X_{k,j}}\left((\widehat{f}_{I-i+1}\cdot...\cdot\widehat{f}_{I-1})^{-1}\right) 
\end{eqnarray}
\begin{eqnarray} & =	&\widehat{\mu}_i\frac{(\widehat{f}_{I-i+1}\cdot...\cdot\widehat{f}_{I-1})\sum_{p=k}^{i-1}\left(\left(\frac{1}{\sum_{q=1}^{p}C_{q,I-p+1}}\right)\textbf{1}_{\{j\leq I-p+1\}}-\left(\frac{1}{\sum_{q=1}^{p}C_{q,I-p}}\right)\textbf{1}_{\{j\leq I-p\}}\right)}{(\widehat{f}_{I-i+1}\cdot...\cdot\widehat{f}_{I-1})^2} \\ 
& =	& \widehat{\mu}_i\frac{\sum_{p=k}^{i-1}\left(\left(\frac{1}{\sum_{q=1}^{p}C_{q,I-p+1}}\right)\textbf{1}_{\{j\leq I-p+1\}}-\left(\frac{1}{\sum_{q=1}^{p}C_{q,I-p}}\right)\textbf{1}_{\{j\leq I-p\}}\right)}{(\widehat{f}_{I-i+1}\cdot...\cdot\widehat{f}_{I-1})} 
\end{eqnarray}
The BF impact function for individual accident year reserves is given as \begin{equation}
\text{IF}_{k,j}(\widehat{R}^{BF}_i)=\begin{cases}
0 & \text{if } k>i-1\\
\widehat{\mu}_i\frac{\sum_{p=k}^{i-1}\left(\left(\frac{1}{\sum_{q=1}^{p}C_{q,I-p+1}}\right)\textbf{1}_{\{j\leq I-p+1\}}-\left(\frac{1}{\sum_{q=1}^{p}C_{q,I-p}}\right)\textbf{1}_{\{j\leq I-p\}}\right)}{(\widehat{f}_{I-i+1}\cdot...\cdot\widehat{f}_{I-1})} & \text{if } k\leq i-1
\end{cases}
\end{equation}
We can formalise the relationship between the magnitudes of $\text{IF}_{k,j}(\widehat{R}_i)$ and $\text{IF}_{k,j}(\widehat{R}_i^{BF})$ for $k\leq i-1$ in terms of the relationship between $\widehat{C}_{i,I}$ and $\widehat{\mu}_i$. We have \begin{equation}
\text{IF}_{k,j}(\widehat{R}_i^{BF})=\widehat{\mu}_i\frac{\sum_{p=k}^{i-1}\left(\left(\frac{1}{\sum_{q=1}^{p}C_{q,I-p+1}}\right)\textbf{1}_{\{j\leq I-p+1\}}-\left(\frac{1}{\sum_{q=1}^{p}C_{q,I-p}}\right)\textbf{1}_{\{j\leq I-p\}}\right)}{(\widehat{f}_{I-i+1}\cdot...\cdot\widehat{f}_{I-1})}
\end{equation} 
and \begin{equation}
\text{IF}_{k,j}(\widehat{R}_i)=\widehat{C}_{i,I}\sum_{p=k}^{i-1}\left(\left(\frac{1}{\sum_{q=1}^{p}C_{q,I-p+1}}\right)\textbf{1}_{\{j\leq I-p+1\}}-\left(\frac{1}{\sum_{q=1}^{p}C_{q,I-p}}\right)\textbf{1}_{\{j\leq I-p\}}\right)
\end{equation}
Hence \begin{eqnarray}
|\text{IF}_{k,j}(\widehat{R}_i^{BF})|<|\text{IF}_{k,j}(\widehat{R}_i)| \Longleftrightarrow
\frac{\widehat{\mu}_i}{(\widehat{f}_{I-i+1}\cdot...\cdot\widehat{f}_{I-1})}<\widehat{C}_{i,I}
\end{eqnarray}
Under the reasonable assumptions that \begin{equation}\label{assumptions}\widehat{\mu}_i,\ (\widehat{f}_{I-i+1}\cdot...\cdot\widehat{f}_{I-1}) \quad \text{and } \quad \widehat{C}_{i,I}>0\end{equation} 
Note that we have focused on the magnitude of the impacts rather than their sign. This is in line with our focus on robustness in that we are concerned with how much each incremental claim can impact reserve estimates. Furthermore, the two impact functions ($\text{IF}_{k,j}(
\widehat{R}_i)$ and $\text{IF}_{k,j}(
\widehat{R}_i^{BF})$) will have the same sign under the assumptions given in \eqref{assumptions}. For completeness the following results are given. If $\sum_{p=k}^{i-1}\left(\left(\frac{1}{\sum_{q=1}^{p}C_{q,I-p+1}}\right)\textbf{1}_{\{j\leq I-p+1\}}-\left(\frac{1}{\sum_{q=1}^{p}C_{q,I-p}}\right)\textbf{1}_{\{j\leq I-p\}}\right)>0$ then 
\begin{equation}
\text{IF}_{k,j}(\widehat{R}_i^{BF})<\text{IF}_{k,j}(\widehat{R}_i) \Longleftrightarrow
\frac{\widehat{\mu}_i}{(\widehat{f}_{I-i+1}\cdot...\cdot\widehat{f}_{I-1})}<\widehat{C}_{i,I}
\end{equation} 
If  $\sum_{p=k}^{i-1}\left(\left(\frac{1}{\sum_{q=1}^{p}C_{q,I-p+1}}\right)\textbf{1}_{\{j\leq I-p+1\}}-\left(\frac{1}{\sum_{q=1}^{p}C_{q,I-p}}\right)\textbf{1}_{\{j\leq I-p\}}\right)<0$ then 
\begin{equation}
\text{IF}_{k,j}(\widehat{R}_i^{BF})<\text{IF}_{k,j}(\widehat{R}_i) \Longleftrightarrow
\frac{\widehat{\mu}_i}{(\widehat{f}_{I-i+1}\cdot...\cdot\widehat{f}_{I-1})}>\widehat{C}_{i,I}
\end{equation} 
\subsection{Independence of reserve impact and $X_{k,j}$ for last diagonal} \label{independence}
We have \begin{eqnarray}
	\text{IF}_{k,j}(\widehat{R}_i)
	& =	& \begin{cases}\label{influence}
		0, &\text{if } k>i \\
		\frac{\widehat{R}_i}{C_{i,I-i+1}},& \text{if } k=i \\ 
		\widehat{C}_{i,I}\sum_{p=k}^{i-1}((\frac{1}{\sum_{q=1}^{p}C_{q,I-p+1}})\textbf{1}_{\{j\leq I-p+1\}}-(\frac{1}{\sum_{q=1}^{p}C_{q,I-p}})\textbf{1}_{\{j\leq I-p\}}), &\text{if } k\leq i-1 
	\end{cases}
\end{eqnarray}
The independence is clear for $k>i$. For $k=i$ we have \begin{eqnarray}\label{ki}
	\text{IF}_{k,j}(\widehat{R}_i)= \frac{\widehat{R}_i}{C_{i,I-i+1}}= \frac{C_{i,I-i+1}\left(\prod_{s=I-i+1}^{I-1}\widehat{f}_s-1\right)}{C_{i,I-i+1}}= \prod_{s=I-i+1}^{I-1}\widehat{f}_s-1
\end{eqnarray}
And \begin{eqnarray}
	\widehat{f}_s=\frac{\sum_{q=1}^{I-s}C_{q,s+1}}{\sum_{q=1}^{I-s}C_{q,s}},\ 1\leq j\leq I-1
\end{eqnarray}
With these development factors, the latest accident year that is considered is $I-I+i-1=i-1$ for $f_{I-i+1}$ and hence by definition this is independent of $X_{k,j}$ where $k=i$. \\
Now for $k\leq i-1$ we have \begin{equation}
	\text{IF}_{k,j}=	\widehat{C}_{i,I}\sum_{p=k}^{i-1}\left(\left(\frac{1}{\sum_{q=1}^{p}C_{q,I-p+1}}\right)\textbf{1}_{\{j\leq I-p+1\}}-\left(\frac{1}{\sum_{q=1}^{p}C_{q,I-p}}\right)\textbf{1}_{\{j\leq I-p\}}\right)
\end{equation}
Given we are looking at observations on the last diagonal (i.e $k+j=I+1$) we have that $j=I-k+1$. Furthermore we have that $\mathbb{D}=\{1,...,i-1\}$ and we are summing over $p\in\mathbb{D}$ for $p\geq k$. Let us first consider the case where $k=i-1$. We have that \begin{eqnarray}
	\text{IF}_{k,j}(\widehat{R}_i)=\widehat{C}_{i,I}\left(\frac{1}{\sum_{q=1}^{i-1}C_{q,I-i+2}}\right)& =	& C_{i,I-i+1}\widehat{f}_{I-i+1}\cdots \widehat{f}_{I-1}\left(\frac{1}{\sum_{q=1}^{i-1}C_{q,I-i+2}}\right) \\
	& =	& C_{i,I-i+1}\left(\frac{1}{\sum_{q=1}^{i-1}C_{q,I-i+1}}\right)\widehat{f}_{I-i+2}\cdots \widehat{f}_{I-1}\label{dev}
\end{eqnarray}
Since we have that $j=I-k+1=I-i+2$ and hence the second indicator function is not satisfied and $\widehat{f}_{I-i+1}=\frac{\sum_{q=1}^{i-1}C_{q,I-i+2}}{\sum_{q=1}^{i-1}C_{q,I-i+1}}$ such that \begin{eqnarray}
	\widehat{f}_{I-i+1}\left(\frac{1}{\sum_{q=1}^{i-1}C_{q,I-i+2}}\right)=\frac{\sum_{q=1}^{i-1}C_{q,I-i+2}}{\sum_{q=1}^{i-1}C_{q,I-i+1}}\left(\frac{1}{\sum_{q=1}^{i-1}C_{q,I-i+2}}\right)
	= \frac{1}{\sum_{q=1}^{i-1}C_{q,I-i+1}}
\end{eqnarray}
    Since $j=I-i+1$ this term is independent of $X_{k,j}=X_{i-1,I-i+2}$. Furthermore for the other development factors in equation \eqref{dev} (i.e $\widehat{f}_{I-i+2}\cdots \widehat{f}_{I-1}$) the latest development factor that is considered is $i-2$ and hence we can see that all members of this expression are independent of $X_{k,j}$. Now as we decrease $k$ (i.e. move from $k=i-1$ to $k=i-2,...,1$) we will be considering more cases of $p\in\mathbb{D}$ since $p\geq k$ will be satisfied however note also that as we decrease $k$ we will be increasing $j$ since $j=I-k+1$ such that for $k=i-2$ we have $j=I-i+3$ which means that both the indicator functions for the case when $p=i-1$ will not be satisfied and only the first will be satisfied for the case when $p=i-2$. Hence we have that for $k=i-2$ \begin{eqnarray}
	\text{IF}_{k,j}(\widehat{R}_i)	& =	& C_{i,I-i+1}\widehat{f}_{I-i+1}\cdots\widehat{f}_{I-1}\left(\frac{1}{\sum_{q=1}^{i-2}C_{q,I-i+3}}\right)\\ 
	& =	& C_{I-i+i}\widehat{f}_{I-i+1}\frac{\sum_{q=1}^{i-2}C_{q,I-i+3}}{\sum_{q=1}^{i-2}C_{q,I-i+2}}\widehat{f}_{I-i+3}\cdots\widehat{f}_{I-1}\left(\frac{1}{\sum_{q=1}^{i-2}C_{q,I-i+3}}\right)\\
	& =	& C_{I-i+i}\widehat{f}_{I-i+1}\frac{1}{\sum_{q=1}^{i-2}C_{q,I-i+2}}\widehat{f}_{I-i+3}\cdots\widehat{f}_{I-1}
\end{eqnarray}
This pattern continues along the last diagonal such that for $k+j=I+1$ and $k\leq i-1$ we have that \begin{eqnarray}\label{independent}
	\text{IF}_{k,j}(X_{k,j})=C_{I-i+1}\left(\frac{1}{\sum_{q=1}^{k}C_{q,I-k}}\right)\prod_{\{s=I-i+1|s\neq I-k\}}^{I-1}\widehat{f}_s
\end{eqnarray}
Note that equation \eqref{independent} is independent of $X_{k,j}$. Hence we have shown that for all values of $k+j=I+1$ (i.e. the last diagonal of the loss triangle) that the impact function for accident year reserves ($\widehat{R}_i$) is independent of the incremental claim itself. This means that the rate of change of reserves with respect to this value is constant such that if the value of this cell were to change by $x$, we would see reserves change by an amount equal to $\text{IF}_{k,j}(\widehat{R}_i)\cdot x$. This result holds for total reserves $\widehat{R}$ too, since $\text{IF}_{k,j}(\widehat{R})=\sum_{i=2}^{I}\text{IF}_{k,j}(\widehat{R}_i)$ which is simply a sum of terms independent of $X_{k,j}$ and hence is also independent of $X_{k,j}$.
\subsection{Homogeneity of Reserves}\label{homogeneity} 
We first show that $\widehat{R}_i$ is homogeneous of order one and it easily follows for $\widehat{R}$. If a function $f(.)$ is homogeneous of order $q$ then $f(t\cdot\textbf{x}) =t^q\cdot f(\textbf{x})$. In this case we have that $\textbf{x}=(X_{1,1},X_{1,2},...,X_{I,1})$. Note for $\ 1\leq j\leq I-1$
\begin{eqnarray}
	\widehat{f}_j=\frac{\sum_{q=1}^{I-s}C_{q,s+1}}{\sum_{q=1}^{I-s}C_{q,s}}
	& =	& \frac{C_{1,s+1}+\cdots+C_{I-s,s+1}}{C_{1,s}+\cdots +C_{I-s,s}}\\
	& =	& \frac{X_{1,1}+\cdots +X_{1,s+1}+\cdots +X_{I-s,1}+\cdots+X_{I-s,s+1}}{X_{1,1}+\cdots +X_{1,s}+\cdots +X_{I-s,1}+\cdots+X_{I-s,s}}
\end{eqnarray}
Now we can see for $g_j(\textbf{x})=\widehat{f}_j$ that $g_j(t\cdot\textbf{x})= g_j(\textbf{x})$ such that estimated  development factors $\widehat{f}_s$ are homogeneous of order zero. Now, reserves are given by and \begin{eqnarray}
	f_i(\textbf{x}) & =	&\widehat{R}_i= C_{i,I-i+1}(\widehat{f}_{I-i+1}\cdots\widehat{f}_{I-1}-1)=(X_{i,1}+\cdots +X_{i,I-i+1})(\widehat{f}_{I-i+1}\cdots\widehat{f}_{I-1}-1) 
\end{eqnarray} 
Such that \begin{eqnarray}
	f_i(t\cdot\textbf{x})=(t\cdot X_{i,1}+\cdots +t\cdot X_{i,I-i+1})(\widehat{f}_{I-i+1}\cdots\widehat{f}_{I-1}-1)=t\cdot f_i(\textbf{x})
\end{eqnarray}
This completes the proof for individual accident year reserves $R_i$. 
Now denote $f(\textbf{x})=\widehat{R}$ such that \begin{eqnarray}
	f(\textbf{x})= \sum_i f_i(\textbf{x})
	\quad \text{and } \quad 
	f(t\cdot\textbf{x}) =\sum_if_i(t\cdot\textbf{x})=\sum_itf_i(\textbf{x})
	=t\sum_if_i(\textbf{x})
\end{eqnarray}
This completes the proof for total reserves $\widehat{R}$.\\[1ex] Now since both $\widehat{R}_i$ and $\widehat{R}$ are homogeneous functions of order 1 we have by Euler's homogeneous function theorem that \begin{eqnarray}
	\widehat{R}_i=\sum_{\{k+j\leq I+1\}}\frac{\partial \widehat{R}_i}{\partial X_{k,j}}\cdot X_{k,j}=\sum_{\{k+j\leq I+1\}}\text{IF}_{k,j}(\widehat{R}_i)\cdot X_{k,j}
\end{eqnarray}
And 
\begin{eqnarray}
	\widehat{R}=\sum_{\{k+j\leq I+1\}}\frac{\partial \widehat{R}}{\partial X_{k,j}}\cdot X_{k,j}=\sum_{\{k+j\leq I+1\}}\text{IF}_{k,j}(\widehat{R})\cdot X_{k,j}
\end{eqnarray}
This implies that $\text{IF}_{k,j}(\widehat{R})\cdot X_{k,j}$ gives the marginal contribution of $X_{k,j}$ to reserves $\widehat{R}$. 
\subsection{$\text{IF}_{k,j}(\text{mse}(\widehat{R}_i))$} \label{mseriproof}
\begin{eqnarray}\label{IF}
	\text{IF}_{k,j}(\text{mse}(\widehat{R}_i))	& =	& \frac{\partial}{\partial X_{k,j}}\left(C_{i,I-i+1}\sum_{j=I-i+1}^{I-1}(f_{I-i+1}\cdots f_{j-1}\sigma^2_jf^2_{j+1}\cdots f^2_{I-1})\right.\\ 
	& +	& \left.C_{i,I-i+1}^2(f_{I-i+1}\cdots f_{I-1}-\widehat{f}_{I-i+1}\cdots \widehat{f}_{I-1})^2\right) \nonumber \\ 
	& =	&
	\begin{split} \sum_{j=I-i+1}^{I-1}(f_{I-i+1}\cdot...\cdot f_{j-1}\sigma_j^2f_{j+1}^2\cdot...\cdot f^2_{I-1})\frac{\partial C_{i,I-i+1}}{\partial X_{k,j}}+\\\frac{\partial (C_{i,I-i+1}(f_{I-i+1}\cdots f_{I-1}-\widehat{f}_{I-i+1}\cdots \widehat{f}_{I-1}))^2}{\partial X_{k,j}}
	\end{split}
\end{eqnarray}
We have \begin{eqnarray}
	\frac{\partial C_{i,I-i+1}}{\partial X_{k,j}}=
	\begin{cases}
		1, &\text{if } k=i \\
		0, &\text{otherwise}
	\end{cases}
\end{eqnarray}
Further
\begin{equation}
	\frac{\partial \left(C_{i,I-i+1}(f_{I-i+1}\cdot...\cdot f_{I-1}-\widehat{f}_{I-i+1}\cdot...\cdot\widehat{f}_{I-1})\right)^2}{\partial X_{k,j}}
\end{equation}
\begin{equation}=2C_{i,I-i+1}(f_{I-i+1}\cdot...\cdot f_{I-1}- \widehat{f}_{I-i+1}\cdot...\cdot\widehat{f}_{I-1})\frac{\partial}{\partial X_{k,j}}C_{i,I-i+1}(f_{I-i+1}\cdot...\cdot f_{I-1}- \widehat{f}_{I-i+1}\cdot...\cdot\widehat{f}_{I-1}) \nonumber
\end{equation}
\begin{small}
	\begin{eqnarray}
		=\begin{cases}
			0 & \text{if } k>i \\
			2C_{i,I-i+1}(f_{I-i+1}\cdot...\cdot f_{I-1}-\widehat{f}_{I-i+1}\cdot...\cdot\widehat{f}_{I-1})^2 & \text{if } k=i \\[1ex]
			-2C_{i,I-i+1}(f_{I-i+1}\cdot...\cdot f_{I-1}-\widehat{f}_{I-i+1}\cdot...\cdot\widehat{f}_{I-1})\widehat{C}_{i,I}\sum_{p=k}^{i-1}\left(\left(\frac{1}{\sum_{i=1}^{p}C_{i,I-p+1}}\right)\textbf{1}_{\{j\leq I-p+1\}}\right.\\\left.-\left(\frac{1}{\sum_{i=1}^{p}C_{i,I-p}}\right)\textbf{1}_{\{j\leq I-p\}}\right) & \text{if } k\leq i-1\nonumber
		\end{cases}
	\end{eqnarray}
\end{small}
\noindent \\[1ex] Note that the result for $k\leq i$ follows from what was previously calculated for the impact function for reserves.
From here we can calculate the impact function for $\text{mse}(\widehat{R}_i)$. It is given by 
\begin{eqnarray}
	\text{IF}_{k,j}(\text{mse}(\widehat{R}_i))=\begin{cases}
		0 &\text{if } k>i\\[1ex]
		\sum_{j=I-i+1}^{I-1}(f_{I-i+1}\cdot...\cdot f_{j-1}\sigma^2_jf_{j+1}^2\cdot...\cdot f_{I-1}^2)+\\2C_{i,I-i+1}(f_{I-i+1}\cdot...\cdot f_{I-1}-\widehat{f}_{I-i+1}\cdot...\cdot\widehat{f}_{I-1})^2 & \text{if } k=i\\[1ex]
		-2C_{i,I-i+1}(f_{I-i+1}\cdot...\cdot f_{I-1}-\widehat{f}_{I-i+1}\cdot...\cdot\widehat{f}_{I-1})\\\widehat{C}_{i,I}\sum_{p=k}^{i-1}\left(\left(\frac{1}{\sum_{i=1}^{p}C_{i,I-p+1}}\right)\textbf{1}_{\{j\leq I-p+1\}}-\left(\frac{1}{\sum_{i=1}^{p}C_{i,I-p}}\right)\textbf{1}_{\{j\leq I-p\}}\right) & \text{if } k\leq i-1 
	\end{cases}
\end{eqnarray}
Note that for the case $k=i$, if we replace the $f_s$ with $\widehat{f}_s$ then the second term be equal to zero. Similarly, for the case when $k\leq i-1$ if we made this substitution the whole term would be equal to zero. Instead we use the estimate of $(f_{I-i+1}\cdot...\cdot f_{I-1}-\widehat{f}_{I-i+1}\cdot...\cdot \widehat{f}_{I-1})^2$ given in Mack (1993) such that. \begin{eqnarray}
	(f_{I-i+1}\cdot...\cdot f_{I-1}-\widehat{f}_{I-i+1}\cdot...\cdot \widehat{f}_{I-1})^2\simeq(\widehat{f}_{I-i+1}\cdot...\cdot \widehat{f}_{I-1})^2\sum_{s=I-i+1}^{I-1}\frac{\widehat{\sigma}^2_s/\widehat{f}_s^2}{\sum_{i=1}^{I-s}C_{i,s}}
\end{eqnarray}
Hence the impact function is given by 
\begin{eqnarray}
	\text{IF}_{k,j}(mse(\widehat{R}_i))=\begin{cases}
		0 &\text{if } k>i\\[1ex]
		\sum_{j=I-i+1}^{I-1}(f_{I-i+1}\cdot...\cdot f_{j-1}\sigma^2_jf_{j+1}^2\cdot...\cdot f_{I-1}^2)+\\2C_{i,I-i+1}(\widehat{f}_{I-i+1}\cdot...\cdot \widehat{f}_{I-1})^2\sum_{s=I-i+1}^{I-1}\frac{\widehat{\sigma}^2_s/\widehat{f}_s^2}{\sum_{i=1}^{I-s}C_{i,s}} & \text{if } k=i\\[1ex]
		-2C_{i,I-i+1}(\widehat{f}_{I-i+1}\cdot...\cdot \widehat{f}_{I-1})\sqrt{\sum_{s=I-i+1}^{I-1}\frac{\widehat{\sigma}^2_s/\widehat{f}_s^2}{\sum_{i=1}^{I-s}C_{i,s}}}\\\widehat{C}_{i,I}\sum_{p=k}^{i-1}\left(\left(\frac{1}{\sum_{i=1}^{p}C_{i,I-p+1}}\right)\textbf{1}_{\{j\leq I-p+1\}}-\left(\frac{1}{\sum_{i=1}^{p}C_{i,I-p}}\right)\textbf{1}_{\{j\leq I-p\}}\right) & \text{if } k\leq i-1 
	\end{cases}
\end{eqnarray}
\subsection{Sign change for IF$_{k,j}($mse$(\widehat{R}_i))$}\label{signchange}
For $k=i-1$ We have \begin{equation}
	\begin{split}
		\text{IF}_{k,j}(\text{mse}(\widehat{R}_i))=
		-2C_{i,I-i+1}(\widehat{f}_{I-i+1}\cdot...\cdot \widehat{f}_{I-1})\sqrt{\sum_{s=I-i+1}^{I-1}\frac{\widehat{\sigma}^2_s/\widehat{f}_s^2}{\sum_{i=1}^{I-s}C_{i,s}}}\\\widehat{C}_{i,I}\sum_{p=k}^{i-1}\left(\left(\frac{1}{\sum_{i=1}^{i-1}C_{i,I-i+2}}\right)\textbf{1}_{\{j\leq I-i+1\}}-\left(\frac{1}{\sum_{i=1}^{i-1}C_{i,I-i+1}}\right)\textbf{1}_{\{j\leq I-i+1\}}\right)  
	\end{split}
\end{equation}
Now for $j\leq I-i+1$, {\footnotesize \begin{eqnarray}
		\begin{split}
			\left(\frac{1}{\sum_{q=1}^{i-1}C_{q,I-i+2}}\right)\textbf{1}_{\{j\leq I-i+2\}}-\left(\frac{1}{\sum_{q=1}^{i-1}C_{q,I-i+1}}\right)\textbf{1}_{\{j\leq I-i+1\}})& =	&\left(\frac{1}{\sum_{q=1}^{i-1}C_{q,I-i+2}}\right)-\left(\frac{1}{\sum_{q=1}^{i-1}C_{q,I-i+1}}\right)\\& < 0	&
		\end{split}
	\end{eqnarray}}however this term becomes  $(\frac{1}{\sum_{q=1}^{i-1}C_{q,I-i+2}})>0$ for $j=I-i+2$.  \\ 
	Hence we see the sign of the impact change from positive for $j\leq I-i+1$ to negative for $j=I-i+2$.
\subsection{$\text{IF}_{k,j}(\text{mse}(\widehat{R}))$} 
	\begin{equation}
		\text{mse}(\widehat{R})=\sum_{i=2}^{I}\left\{\left(\text{s.e.}(\widehat{R}_i)\right)^2+\widehat{C}_{i,I}\left(\sum_{q=i+1}^{I}\widehat{C}_{q,I}\right)\left(\sum_{r=I-i+1}^{I-1}\frac{2\sigma_r^2/\widehat{f}_r^2}{\sum_{n=1}^{I-r}C_{n,r}}\right)\right\}
	\end{equation}
	The impact function for $(\text{s.e.}(\widehat{R}_i))^2$ terms will be the same as is given by equation \eqref{ifmse}. Hence our focus turns to the second term inside the summand, i.e. $\widehat{C}_{i,I}\left(\sum_{q=i+1}^{I}\widehat{C}_{q,I}\right)\left(\sum_{r=I-i+1}^{I-1}\frac{2\sigma_r^2/\widehat{f}_r^2}{\sum_{n=1}^{I-r}C_{n,r}}\right)$. To find the impact function for this term we make use of the product rule. In this vein let $u=\widehat{C}_{i,I}(\sum_{q=i+1}^{I}\widehat{C}_{q,I})$ and $v=\sum_{r=I-i+1}^{I-1}\frac{2\sigma_r^2/\widehat{f}_r^2}{\sum_{n=1}^{I-r}C_{n,r}}$. For $u$ we have that \begin{equation}
		\frac{\partial \widehat{C}_{i,I}}{\partial X_{k,j}}=\text{IF}_{k,j}(\widehat{R}_i)+\frac{\partial C_{i,I-i+1}}{\partial X_{k,j}}
	\end{equation}
	Further 
	\begin{equation}
		\frac{\partial}{\partial X_{k,j}} \left(\sum_{q=i+1}^{I}\widehat{C}_{q,I}\right)=\sum_{q=i+1}^{I}\left(\text{IF}_{k,j}(\widehat{R}_q)+\frac{\partial C_{q,I-q+1}}{\partial X_{k,j}}\right)
	\end{equation}
	Hence we have \begin{equation}
		u'=\widehat{C}_{i,I}\left(\sum_{q=i+1}^{I}\left(\text{IF}_{k,j}(\widehat{R}_q)+\frac{\partial C_{q,I-q+1}}{\partial X_{k,j}}\right)\right)+\left(\sum_{q=i+1}^{I}\widehat{C}_{q,I}\right)\left(\text{IF}_{k,j}(\widehat{R}_i)+\frac{\partial C_{i,I-i+1}}{\partial X_{k,j}}\right)
	\end{equation}
	Now we focus on \begin{equation}v=	\sum_{r=I-i+1}^{I-1}\frac{2\sigma_r^2/\widehat{f}_r^2}{\sum_{n=1}^{I-r}C_{n,r}}= \sum_{r=I-i+1}^{I-1}2\sigma_r^2\left(\sum_{n=1}^{I-r}C_{n,r}\widehat{f}_r^2\right)^{-1}
	\end{equation}
	Hence \begin{eqnarray}
		\frac{\partial}{\partial X_{k,j}}\left( \sum_{r=I-i+1}^{I-1}\frac{2\sigma_r^2/\widehat{f}_r^2}{\sum_{n=1}^{I-r}C_{n,r}}\right)	& =	& \sum_{r=I-i+1}^{I-1}\frac{-2\sigma_r^2\frac{\partial }{\partial X_{k,j}}\sum_{n=1}^{I-r}C_{n,r}\widehat{f}_r^2}{\left(\sum_{n=1}^{I-r}C_{n,r}\widehat{f}_r^2\right)^2}
	\end{eqnarray}
	Now \begin{eqnarray}
		\frac{\partial}{\partial X_{k,j}}\left(\sum_{n=1}^{I-r}C_{n,r}\widehat{f}_r^2\right)=\sum_{n=1}^{I-r}\widehat{f}_r^2C_{n,r}\left(\frac{\partial\ln  C_{n,r}}{\partial X_{k,j}}+\frac{2\partial\ln  \widehat{f}_r}{\partial X_{k,j}}\right)
	\end{eqnarray}
	Hence we have \begin{equation}
		v' =\sum_{r=I-i+1}^{I-1}\frac{-2\sigma_r^2\frac{\partial }{\partial X_{k,j}}\sum_{n=1}^{I-r}C_{n,r}\widehat{f}_r^2}{\left(\sum_{n=1}^{I-r}C_{n,r}\widehat{f}_r^2\right)^2}= \sum_{r=I-i+1}^{I-1}\frac{-2\sigma_r^2\sum_{n=1}^{I-r}\widehat{f}_r^2C_{n,r}\left(\frac{\partial\ln  C_{n,r}}{\partial X_{k,j}}+\frac{2\partial\ln  \widehat{f}_r}{\partial X_{k,j}}\right)}{\left(\sum_{n=1}^{I-r}C_{n,r}\widehat{f}_r^2\right)^2}
	\end{equation}
	Hence the impact function for the mse of the estimated total reserves is given by 
	\begin{footnotesize}
		\begin{eqnarray}
			\begin{split}
				\text{IF}_{k,j}\left(\text{mse}(\widehat{R})\right)=\sum_{i=2}^{I}\left\{\widehat{\text{IF}}_{k,j}\left(\text{mse}(\widehat{R}_i)\right)+\widehat{C}_{i,I}\left(\sum_{q=i+1}^{I}\widehat{C}_{q,I}\right)\sum_{r=I-i+1}^{I-1}\frac{-2\sigma_r^2\sum_{n=1}^{I-r}\widehat{f}_r^2C_{n,r}\left(\frac{\partial\ln  C_{n,r}}{\partial X_{k,j}}+\frac{2\partial\ln  \widehat{f}_r}{\partial X_{k,j}}\right)}{\left(\sum_{n=1}^{I-r}C_{n,r}\widehat{f}_r^2\right)^2}+\right.\\\left.\left(\sum_{r=I-i+1}^{I-1}\frac{2\sigma_r^2/\widehat{f}_r^2}{\sum_{n=1}^{I-r}C_{n,r}}\right)\left(\widehat{C}_{i,I}\left(\sum_{q=i+1}^{I}\left(\text{IF}_{k,j}(\widehat{R}_q)+\frac{\partial C_{q,I-q+1}}{\partial X_{k,j}}\right)\right)+\left(\sum_{q=i+1}^{I}\widehat{C}_{q,I}\right)(\text{IF}_{k,j}\left(\widehat{R}_i)+\frac{\partial C_{i,I-i+1}}{\partial X_{k,j}}\right)\right)\right\}
			\end{split}
		\end{eqnarray}
	\end{footnotesize}
\subsection{$\text{IF}_{k,j}(F^{-1}_R(q))$}\label{quantilesproof}
	We assume that total reserves $(R)$ follow a lognormal distribution with  \begin{equation} E[R]=\widehat{R}=e^{\mu+\frac{1}{2}\sigma^2} \quad \text{and} \quad  \text{Var}(R)=\text{mse}(\widehat{R})=e^{2\mu+\sigma^2}(e^{\sigma^2}-1)
	\end{equation}
	Such that \begin{equation}
		R\sim LN(\mu,\sigma^2)
		\quad \text{and } \quad
		F_R^{-1}(q)=e^{\mu+\sigma\Phi^{-1}(q)}
	\end{equation}
	Where $\Phi(.)$ is the cumulative distribution function of the standard normal distribution. Firstly, we must solve for $\mu$ and $\sigma^2$ in terms of $\widehat{R}$ and mse$(\widehat{R})$. Upon rearrangement we have \begin{equation}
		\sigma^2=\ln \left(1+\frac{\text{mse}(\widehat{R})}{\widehat{{R}^2}}\right)
	\end{equation}
	\begin{equation}
		\mu	 =	\ln (\widehat{R})-\frac{1}{2}\sigma^2 
		=\ln (\widehat{R})-\frac{1}{2}\ln \left(1+\frac{\text{mse}(\widehat{R})}{\widehat{{R}^2}}\right)
	\end{equation}
	Hence the quantiles of the reserve distribution are given by \begin{equation}
		F^{-1}_R(q)=\exp\left[\ln (\widehat{R})-\frac{1}{2}\ln \left(1+\frac{\text{mse}(\widehat{R})}{\widehat{{R}^2}}\right)+\sqrt{\ln \left(1+\frac{\text{mse}(\widehat{R})}{\widehat{{R}^2}}\right)}\Phi^{-1}(q)\right]
	\end{equation}
	We can now find the impact that each incremental claim is having on estimated reserve quantiles. \begin{eqnarray}
		\frac{\partial}{\partial X_{k,j}}F_R^{-1}(q)	& =	&\frac{\partial}{\partial X_{k,j}}\exp\left[\ln (\widehat{R})-\frac{1}{2}\ln \left(1+\frac{\text{mse}(\widehat{R})}{\widehat{{R}^2}}\right)+\sqrt{\ln \left(1+\frac{\text{mse}(\widehat{R})}{\widehat{{R}^2}}\right)}\Phi^{-1}(q)\right] \nonumber
	\end{eqnarray}
	\begin{equation}
		\begin{split}
			=\frac{\partial}{\partial X_{k,j}}\left(\ln (\widehat{R})-\frac{1}{2}\ln \left(1+\frac{\text{mse}(\widehat{R})}{\widehat{{R}^2}}\right)+\sqrt{\ln \left(1+\frac{\text{mse}(\widehat{R})}{\widehat{{R}^2}}\right)}\Phi^{-1}(q)\right)\\\cdot\exp\left[\ln (\widehat{R})-\frac{1}{2}\ln \left(1+\frac{\text{mse}(\widehat{R})}{\widehat{{R}^2}}\right)+\sqrt{\ln \left(1+\frac{\text{mse}(\widehat{R})}{\widehat{{R}^2}}\right)}\Phi^{-1}(q)\right] 
		\end{split}
	\end{equation}
	We have
	$\frac{\partial}{\partial X_{k,j}}\ln (\widehat{R})=\frac{\text{IF}_{k,j}(\widehat{R})}{\widehat{R}}$
	and \begin{eqnarray}
		\frac{\partial}{\partial X_{k,j}}\left(-\frac{1}{2}\ln \left(1+\frac{\text{mse}(\widehat{R})}{\widehat{{R}^2}}\right)\right)	& =	& -\frac{1}{2}\frac{\frac{\partial}{\partial X_{k,j}} \left(1+\frac{\text{mse}(\widehat{R})}{\widehat{R}^2}\right)}{\left(1+\frac{\text{mse}(\widehat{R})}{\widehat{R}^2}\right)}
	\end{eqnarray}
	By the quotient rule \begin{eqnarray}
		\frac{\partial}{\partial X_{k,j}}\left(1+\frac{\text{mse}(\widehat{R})}{\widehat{R}^2}\right)=\frac{\partial}{\partial X_{k,j}}\frac{\text{mse}(\widehat{R})}{\widehat{R}^2}=\frac{\text{IF}_{k,j}\left(\text{mse}(\widehat{R})\right)\cdot\widehat{R}^2-2\text{IF}_{k,j}(\widehat{R})\cdot\widehat{R}\cdot\text{mse}(\widehat{R})}{\widehat{R}^4}
	\end{eqnarray}
	Hence \begin{eqnarray}
		\frac{\partial}{\partial X_{k,j}}\left(-\frac{1}{2}\ln \left(1+\frac{\text{mse}(\widehat{R})}{\widehat{{R}^2}}\right)\right)	& =	& -\frac{\text{IF}_{k,j}\left(\text{mse}(\widehat{R})\right)\cdot\widehat{R}-2\text{IF}_{k,j}(\widehat{R})\cdot\text{mse}(\widehat{R}) }{2\widehat{R}\left(\text{mse}(\widehat{R})+\widehat{R}^2\right)}
	\end{eqnarray}
	Such that \begin{eqnarray}
		\frac{\partial}{\partial X_{k,j}}\mu	& =	&\frac{\partial}{\partial X_{k,j}}\left(\ln (\widehat{R})-\frac{1}{2}\ln \left(1+\frac{\text{mse}(\widehat{R})}{\widehat{{R}^2}}\right)\right)\\	& =	& \frac{\text{IF}_{k,j}(\widehat{R})}{\widehat{R}}-\frac{\text{IF}_{k,j}(\text{mse}(\widehat{R}))\cdot\widehat{R}-2\text{IF}_{k,j}(\widehat{R})\cdot\text{mse}(\widehat{R}) }{2\widehat{R}(\text{mse}(\widehat{R})+\widehat{R}^2)} \\
		& =	&\frac{2\cdot \text{IF}_{k,j}(\widehat{R})\cdot\widehat{R}-\text{IF}_{k,j}\left(\text{mse}(\widehat{R})\right)}{2\left(\text{mse}(\widehat{R})+\widehat{R}^2\right)}
	\end{eqnarray}
	Now \begin{eqnarray}
		\frac{\partial}{\partial X_{k,j}}\left(\sqrt{\ln \left(1+\frac{\text{mse}(\widehat{R})}{\widehat{{R}^2}}\right)}\Phi^{-1}(q)\right)	& =	& \Phi^{-1}(q)\frac{\partial}{\partial X_{k,j}}\left(\sqrt{\ln \left(1+\frac{\text{mse}\widehat{R}}{\widehat{R}^2}\right)}\right)\\
		& =	& 	\frac{\Phi^{-1}(q)}{2}\cdot\frac{\frac{\partial}{\partial X_{k,j}}\ln \left(1+\frac{\text{mse}\widehat{R}}{\widehat{R}^2}\right)}{\sqrt{\ln \left(1+\frac{\text{mse}\widehat{R}}{\widehat{R}^2}\right)}}\\
		& =	& \frac{\Phi^{-1}(q)\left(\text{IF}_{k,j}(\text{mse}\left(\widehat{R}\right)\cdot\widehat{R}-2\text{mse}(\widehat{R}))\cdot \text{IF}_{k,j}(\widehat{R})\right)}{2\widehat{R}\left(\text{mse}(\widehat{R})+\widehat{R}^2\right)\sqrt{\ln \left(1+\frac{\text{mse}\widehat{R}}{\widehat{R}^2}\right)}} \nonumber
	\end{eqnarray}
	Finally we have the impact function for the quantiles of the reserve distribution is given by \begin{footnotesize}
		\begin{eqnarray}
			\begin{split}
				\text{IF}_{k,j}\left(F_R^{-1}(q)\right)=\left(\frac{2\cdot \text{IF}_{k,j}(\widehat{R})\cdot\widehat{R}-\text{IF}_{k,j}\left(\text{mse}(\widehat{R})\right)}{2\left(\text{mse}(\widehat{R})+\widehat{R}^2\right)}+\frac{\Phi^{-1}(q)\left(\text{IF}_{k,j}\left(\text{mse}(\widehat{R})\right)\cdot\widehat{R}-2\text{mse}(\widehat{R})\cdot \text{IF}_{k,j}(\widehat{R})\right)}{2\widehat{R}\left(\text{mse}(\widehat{R})+\widehat{R}^2\right)\sqrt{\ln \left(1+\frac{\text{mse}\widehat{R}}{\widehat{R}^2}\right)}}\right)\cdot\\\exp\left[\ln (\widehat{R})-\frac{1}{2}\ln \left(1+\frac{\text{mse}(\widehat{R})}{\widehat{{R}^2}}\right)+\sqrt{\ln \left(1+\frac{\text{mse}(\widehat{R})}{\widehat{{R}^2}}\right)}\Phi^{-1}(q)\right]
			\end{split}
		\end{eqnarray}
	\end{footnotesize}
\newpage
\section{Multivariate Chain-Ladder}\label{unidata}
We briefly present the multivariate chain-ladder technique put forward by \citet*{MeWu08}. In the following outline, $N$ represents the number of aggregate run-off triangles that are being simultaneously considered and $n$ represents a specific triangle (e.g. line of business) under consideration. %Denote the collection of observations we currently have for triangle $n$ by $\mathbb{B}^{(n)}=\{C_{i,j}^{(n)};i+j\leq I+1\}$. Then the total observations we currently have (i.e. across all triangles) is given by $\mathbb{B}^N=\cup_{n=1}^{N}\mathbb{B}^{(n)}$.%
We have the following matrix operators
\begin{equation*}
\textbf{D}(a) = \left( \begin{array}{ccc}
a_1 & 0 & 0\\
0 & \ddots & 0\\
0 & 0 & a_n \end{array} \right) \quad \text{and } \quad \textbf{D}(a)^b  = \left( \begin{array}{ccc}
a_1^b & 0 & 0 \\
0 & \ddots & 0\\
0 & 0 & a_n^b \end{array} \right)
\end{equation*}
Denote by $\textbf{C}_{ij}$ the vector representing the cumulative claims for accident year $i$ and development year $j$. Firstly, it is assumed that $\textbf{C}_{ij}$ are independent for different accident years $i$. It is also assumed that the following constants $\textbf{f}_j=(f_j^{(1)},...,f_j^{(N)})'$  and   $\pmb{\sigma}_j=(\sigma_j^{(1)},...,\sigma_j^{(N)})'$,
and the following independent RVs $(\pmb{\epsilon}_{i,j+1}^{(1)},...,\pmb{\epsilon}_{i,j+1}^{(N)})$,
	exist such that
	\begin{equation}
	\textbf{C}_{i,j+1}=D(\textbf{f}_j)\textbf{C}_{ij}+D(\textbf{C}_{ij})^{\frac{1}{2}}D(\pmb{\epsilon}_{i,j+1})\pmb{\sigma}_j
	\end{equation}
	with 
	$E[\pmb{\epsilon}_{i,j+1}]=\textbf{0}$ such that 
	\begin{equation}
	E[\textbf{C}_{i,j+1}|\textbf{C}_{ij}]=D(\textbf{f}_j)\textbf{C}_{ij}
	\quad \text{and } \quad
	\widehat{\textbf{C}}_{i,j}=\prod_{l=I-i}^{j-1}D(\widehat{\textbf{f}}_l)\cdot\textbf{C}_{i,I-i}.
	\end{equation}
The development factors $\textbf{f}_j$ are estimated in a manner that considers the dependence between the various loss triangles. Note that predicted ultimate claims for a single accident year and for all accident years across all triangles are given by \begin{equation}
\sum_{n=1}^{N}\widehat{C}_{i,J}^{(n)}
\quad \text{and } \quad
\sum_{i=1}^{I}\sum_{n=1}^{N}\widehat{C}_{i,J}^{(n)}.
\end{equation}
\citet*{MeWu08} also provide a measure of the mean squared error of reserves. For greater detail regarding the calculation of this measure as well as the parameter estimation procedure we refer the interested reader to the original paper. 

\end{document}